\documentclass{jfm}
\usepackage{amsmath}
\usepackage{graphicx}
\usepackage{natbib}
\usepackage{color}

\ifCUPmtlplainloaded \else
  \checkfont{eurm10}
  \iffontfound
    \IfFileExists{upmath.sty}
      {\typeout{^^JFound AMS Euler Roman fonts on the system,
                   using the 'upmath' package.^^J}%
       \usepackage{upmath}}
      {\typeout{^^JFound AMS Euler Roman fonts on the system, but you
                   don't seem to have the}%
       \typeout{'upmath' package installed. jfm.cls can take advantage
                 of these fonts,^^Jif you use 'upmath' package.^^J}%
      }
  \else
  \fi
\fi

\ifCUPmtlplainloaded \else
  \checkfont{msam10}
  \iffontfound
    \IfFileExists{amssymb.sty}
      {\typeout{^^JFound AMS Symbol fonts on the system, using the
                'amssymb' package.^^J}%
       \usepackage{amssymb}%
       \let\le=\leqslant  \let\leq=\leqslant
       \let\ge=\geqslant  \let\geq=\geqslant
      }{}
  \fi
\fi

\ifCUPmtlplainloaded \else
  \IfFileExists{amsbsy.sty}
    {\typeout{^^JFound the 'amsbsy' package on the system, using it.^^J}%
     \usepackage{amsbsy}}
    {\providecommand\boldsymbol[1]{\mbox{\boldmath $##1$}}}
\fi

\newcommand\Real{\mbox{Re}} 
\newcommand\Imag{\mbox{Im}} 

\newcommand{\dy}{\partial}

\newcommand{\ex}{\mathrm{e}}

\newcommand{\zi}{{\rm i}}

\newcommand{\grad}{\nabla}

\newcommand{\eb}{\boldsymbol{e}}

\newcommand{\ub}{\boldsymbol{u}}

\newcommand{\Bb}{{\boldsymbol{B}}}

\newcommand{\Ub}{{\boldsymbol{U}}}

\title[Shear instabilities in SWMHD (\today)]{Shear instabilities in
shallow-water magnetohydrodynamics}

\author[J. Mak, S. D. Griffiths and D. W. Hughes (\today)]
{J. Mak\thanks{Email address for correspondence:
julian.c.l.mak@googlemail.com; present address: School of Mathematics,
University of Edinburgh, James Clerk Maxwell Building, The King's Buildings,
Edinburgh, EH9 3FD, UK},
S. D. Griffiths and D. W. Hughes}

\affiliation{Department of Applied Mathematics, University of Leeds, Leeds, LS2
9JT, UK}

\date{\today}
\pagerange{\pageref{firstpage}--\pageref{lastpage}}
\doi{S002211200100456X}

\begin{document}
\maketitle
\label{firstpage}
\begin{abstract}
Within the framework of shallow-water magnetohydrodynamics, we investigate the
linear instability of horizontal shear flows, influenced by an aligned magnetic
field and stratification. Various classical instability results, such as
H{\o}iland's growth rate bound and Howard's semi-circle theorem, are extended to
this shallow-water system for quite general profiles. Two specific
piecewise-constant velocity profiles, the vortex sheet and the rectangular jet,
are studied analytically and asymptotically; it is found that the magnetic field
and stratification (as measured by the Froude number) are generally both
stabilising, but weak instabilities can be found at arbitrarily large Froude
number. Numerical solutions are computed for corresponding smooth velocity
profiles, the hyperbolic-tangent shear layer and the Bickley jet, for a uniform
background field. A generalisation of the long-wave asymptotic analysis of
\cite{DrazinHoward62} is employed in order to understand the instability
characteristics for both profiles. For the shear layer, the mechanism underlying
the primary instability is interpreted in terms of counter-propagating Rossby
waves, thereby allowing an explication of the stabilising effects of the
magnetic field and stratification. 

\end{abstract}

 



\section{Introduction}

The interaction of horizontal shear flows and magnetic fields in stably
stratified layers is central to many problems in astrophysical fluid dynamics
--- involving, for example, planetary interiors, stellar radiative zones and
accretion discs. An important example of such a flow, which has received
considerable attention recently, is that of the solar tachocline (see Hughes,
Rosner \& Weiss \citeyear{Hughes-et-al-Tachocline}). The tachocline, discovered
via helioseismic observations, is a thin layer in the Sun, extending downwards
from the (neutrally stable) base of the convective zone to the (stably
stratified) top of the radiative interior, characterised by radial velocity
shear and also planetary scale horizontal shears, associated with the equator to
pole differential rotation of the Sun. Most models of the solar dynamo invoke
the tachocline as the site for the storage and generation of the Sun's strong,
predominantly toroidal magnetic field.

Here we are interested in the stability of such shear flows, and how this
depends upon the velocity profile, magnetic field strength, and stratification.
Specifically, we consider the linear stability of a steady parallel flow and
aligned magnetic field, both sheared in the horizontal cross-stream direction,
in the inviscid and perfectly conducting limit. In this first study, we consider
the case where there is no background rotation. The nonlinear regime of such
instabilities typically leads to turbulent flows; these may be important for
dynamo action, through some mean-field $\alpha$-effect, and also for the
transport of mass and momentum, which can feed back on the large-scale flow. 

It is possible to examine the stability of such flows in a continuously
stratified three-dimensional setting \citep[e.g.,][]{MiuraPritchett82, Cally03}.
However, here we adopt the alternative approach of considering the dynamics of a
thin fluid layer under the shallow-water approximation, which is valid when the
horizontal length scale of the motion is long compared with the depth of the
fluid layer, as is typically the case in large-scale astrophysical flows. This
leads to a set of two-dimensional partial differential equations, with no
explicit dependence on the vertical co-ordinate, which offers a considerable
mathematical simplification. Such shallow-water equations capture the
fundamental dynamics of density stratification, including gravity waves, and
allow the interaction of stratification with horizontal shear flows and magnetic
fields to be analysed in the simplest possible setting. 

It should be noted that hydrodynamic shallow-water models, which date back to
Laplace, are derived by considering a thin fluid layer of constant density
bounded below by a rigid medium and above by a fluid of negligible inertia
\citep[e.g.,][\S3.1]{Vallis-GFD}. The corresponding reduction for electrically
conducting fluids --- the shallow-water magnetohydrodynamic (SWMHD) equations of
\cite{Gilman00} --- additionally requires the fluid layer to be a perfect
conductor, and to be bounded above and below by perfect conductors. There are
few direct astrophysical analogues for such a configuration. However, we can
borrow an important idea from planetary atmospheric dynamics, where the
hydrodynamic shallow-water equations are widely used to understand waves and
instabilities in a continuously stratified atmospheric layer. This is justified
because there is a formal mathematical analogy between the linearised equations
in the two systems, provided the layer depth in the shallow-water model is taken
to be a so-called equivalent depth \cite[e.g.,][\S6.11]{Gill-GFD}, so that the
shallow-water gravity wave speed (in the horizontal) matches that of (the
fastest) gravity waves in a continuously stratified layer. We have this analogy
in mind throughout this study.  

The SWMHD equations have been studied widely in recent years. They have been
shown to possess a hyperbolic as well as a Hamiltonian structure
\citep{DeSterck01, Dellar02}, and to support wave motions such as
inertia-gravity waves and Alfv\'en waves \citep{Schecter-et-al01,
Zaqarashvili-et-al08, HengSpitkovsky09}. As reviewed by \cite{GilmanCally07},
they have also been used to study the linear instability of shear flows in
spherical geometry, often with modelling tachocline instabilities in mind. These
previous studies considered basic states that were functions only of latitude,
investigating the dependence of the instabilities on the strength and spatial
structure of the magnetic field and on a (reduced) gravity parameter
\citep{GilmanDikpati02, Dikpati-et-al03}.

In contrast to previous investigations of the instabilities of shear flows in
SWMHD, which focused on global instabilities in spherical geometry,
here we concentrate on local instabilities, with the aim of examining
the linear instability problem in a wider context; for this, we consider the
problem in planar geometry. We first derive some growth rate bounds and
stability criteria, valid for general basic states. We then study how the
instability characteristics of prototypical flows are modified by the combined
action of magnetic fields and stratification, which, in isolation, are generally
thought to be stabilising. The corresponding hydrodynamic problem has a long
history, dating back to Rayleigh, and we are able to draw upon
ideas and methods from a substantial literature
\citep[e.g.,][]{DrazinReid-Stability, Vallis-GFD}. 

We start, in \S2, by formulating the linear instability problem for
plane-parallel basic states with the flow and field dependent on the
cross-stream direction. In \S3, we derive extensions of classical growth rate
bounds, semi-circle theorems, stability criteria, and parity results for modal
solutions. In \S4, instabilities of piecewise-constant velocity profiles (the
vortex sheet and the rectangular jet) with a uniform magnetic field are examined
analytically. Analogous smooth profiles (the hyperbolic-tangent shear layer and
the Bickley jet) are studied in \S5, both numerically and asymptotically, via a
generalisation of the long-wave analysis of \cite{DrazinHoward62}. The primary
instability mechanism for shear layers is interpreted in terms of
counter-propagating Rossby waves. The results are discussed in \S6.



\section{Mathematical formulation}


\subsection{Governing equations}

We consider a thin layer of perfectly electrically conducting fluid moving under
the influence of gravity. We use a Cartesian geometry, with horizontal
coordinates $x$ and $y$, and an upwards pointing coordinate $z$. At time $t$,
the fluid, which is taken to be inviscid and of constant density $\rho$, has a
free surface at $z = h(x,y,t)$ and is bounded below by a rigid impermeable
boundary at $z=-H(x,y)$.

We consider motions with a characteristic horizontal length scale $L_0$ that is
long compared with a characteristic layer depth $H_0$. One can then make a
shallow-water reduction in which the vertical momentum balance is taken to be
magneto-hydrostatic, and for which the horizontal velocity $\ub$ and horizontal
magnetic field $\Bb$ are independent of $z$. When the bottom boundary is
perfectly electrically conducting (and is thus a magnetic field line) and the
free surface remains a field line, the magnetic shallow-water equations of
\cite{Gilman00} are obtained. These are an extension of the classical
shallow-water equations of geophysical fluid dynamics.

We use these equations in non-dimensional form. We denote the characteristic
horizontal velocity of the basic state by $U_0$, and the characteristic magnetic
field strength by $B_0$. Non-dimensionalising $x$ and $y$ by $L_0$, $t$ by the
advective time-scale $L_0/U_0$, $H$ by $H_0$, $h$ by $U_0^2/g$ (where $g$ is the
acceleration due to gravity), velocity by $U_0$, and magnetic field by $B_0$,
the SWMHD equations are
\begin{subequations}\label{ch3:basic-equation}
\begin{align}
	\frac{\dy\ub}{\dy{t}}+\ub\cdot\grad\ub & = -\grad h + M^2 \Bb\cdot\grad\Bb, 
	\label{ch3:basic-equation1} \\
	\frac{\dy\Bb}{\dy{t}}+\ub\cdot\grad\Bb & = \Bb\cdot\grad\ub, 
	\label{ch3:basic-equation2} \\
	F^{2}\frac{\dy{h}}{\dy{t}}+\grad\cdot\left((H+F^{2}h)\ub\right) & =0, 
	\label{ch3:basic-equation3}
\end{align}
\end{subequations}
where $F=U_0/\sqrt{gH_0}$ and $M=(B_0/\sqrt{\mu\rho})/U_0$, with $\mu$ being the
permeability of the fluid. In addition to (\ref{ch3:basic-equation}$a$--$c$),
the shallow-water reduction implies
\begin{equation}\label{ch3:basic-equation4}
	\grad\cdot\left((H+F^{2}h)\Bb\right)=0. 
\end{equation}
However, \eqref{ch3:basic-equation4} need not be considered explicitly; if it is
satisfied at some initial time, then (\ref{ch3:basic-equation}$a$--$c$)
guarantee that it remains satisfied for all time. 

The system has two non-dimensional parameters. The Froude number $F$ is the
ratio of the characteristic horizontal velocity of the basic state to the
gravity wave speed $\sqrt{gH_0}$ (and is related to the reduced gravity
parameter $G$ of \cite{GilmanDikpati02} via $G=F^{-2}$). The parameter $M$ is
the ratio of the Alfv\'en wave speed $B_0/\sqrt{\mu \rho}$ to the characteristic
horizontal velocity of the basic state. When $H$ is constant and $F \rightarrow
0$, (\ref{ch3:basic-equation3}) and (\ref{ch3:basic-equation4}) become $\grad
\cdot \ub = 0$ and $\grad \cdot \Bb = 0$ respectively, and we recover the
equations for two-dimensional incompressible magnetohydrodynamics, with $h$
playing the role of pressure. When $M \rightarrow 0$,
\eqref{ch3:basic-equation2} decouples from (\ref{ch3:basic-equation}$a$--$c$),
and we recover the hydrodynamic shallow-water equations; these have a well-known
correspondence with two-dimensional compressible hydrodynamics
\citep[e.g.,][\S3.1]{Vallis-GFD}, which we exploit from time to time. 

As an example of astrophysical parameter values, we estimate $F$ and $M$ in the
solar tachocline, using data from \cite{Gough07}. We set $U_0$ to be the equator
to pole difference in the zonal velocity, implying $U_0 \approx 500 \,{\rm m \,
s^{-1}}$. There is considerable uncertainty in the strength of the magnetic
field in the tachocline \citep{Hughes-et-al-Tachocline}, although a likely range
is $10^3 \, {\rm G} \lesssim B_0 \lesssim 10^5 \, {\rm G}$. Then, taking $\rho =
210\,{\rm kg \, m^{-3}}$, we find $0.01 \lesssim M \lesssim 1$. To estimate $F$,
we must choose a gravity wave speed $\sqrt{gH_0}$ for the layer. One means of
doing this is to take $H_0$ to be the depth of the tachocline and to interpret
$g$ as a reduced gravity, accounting for the fractional density difference of
the overlying fluid, as in \cite{DikpatiGilman01}. However, here we pursue the
analogy between shallow-water flows and those of a continuously stratified layer
with buoyancy frequency $N$ and depth $H_1$, and choose $\sqrt{gH_0}$ to be the
speed of the fastest gravity wave in such a layer, which is $NH_1/\pi$
\citep[][\S6.11]{Gill-GFD}. Taking $H_1 \approx 2 \times 10^7$\,m (i.e.\ $0.03
R_\odot$, where $R_\odot$ is the solar radius) and $N \approx 8 \times 10^{-4}
\, {\rm s}^{-1}$, which are bulk values that might describe a mode spanning the
entire tachocline, gives a gravity wave speed $N H_1 / \pi = \sqrt{gH_0} \approx
5000 \, {\rm m \, s^{-1}}$, corresponding to an equivalent depth $H_0 \approx
50$\,km (taking $g \approx 540 \, {\rm m \, s^{-2}}$). Again taking $U_0 \approx
500 \,{\rm m\, s^{-1}}$, we thus estimate $F \approx 0.1$, although it is clear
that $F$ would be somewhat smaller or larger if one considered motions towards
the top of the radiative zone (with stronger stratification) or towards the base
of the convection zone (with weaker stratification).



\subsection{The linear instability problem}

Above a topography of the form $H=H(y)$, we consider a basic state $h=0$,
$\ub=U(y)\eb_{x}$ and $\Bb=B(y)\eb_{x}$, so that the magnetic field is initially
aligned with the flow. We then consider perturbations in $h$, $\ub=(u,v)$ and
$\Bb=(b_x,b_y)$ to this state of the form
\begin{equation}\label{ch3:expform}
  	\xi(x,y,t)=\Real\{\hat{\xi}(y)\exp \left(\zi\alpha(x-ct)\right) \}, 
\end{equation}
where $\alpha$ is the (real) wavenumber and $c$ is the (complex) phase speed.
Dropping the hatted notation, the linear evolution is described by
\begin{subequations}\label{ch3:linear-equ1}
\begin{align}
	\left(\frac{\dy}{\dy{t}}+U\frac{\dy}{\dy{x}}\right)u + U' v 
	& = -\frac{\dy{}h}{\dy{x}} 
	+ M^2 \left(B\frac{\dy{}b_{x}}{\dy{x}} + B' b_{y} \right),
	\label{ch3:linear-equ1a} \\
	\left(\frac{\dy}{\dy{t}}+U\frac{\dy}{\dy{x}}\right)v 
	& = -\frac{\dy{h}}{\dy{y}} + M^2 B \frac{\dy{}b_{y}}{\dy{x}},
	\label{ch3:linear-equ1b} \\	
	\left(\frac{\dy}{\dy{t}}+U\frac{\dy}{\dy{x}}\right)b_{x}+B'v 
	& = B\frac{\dy{u}}{\dy{x}} + U'b_{y} ,
	\label{ch3:linear-equ1c} \\
 	\left(\frac{\dy}{\dy{t}}+U\frac{\dy}{\dy{x}}\right)b_{y} 
 	& = B\frac{\dy{v}}{\dy{x}},
 	\label{ch3:linear-equ1d} \\
 	F^2 \left(\frac{\dy}{\dy{t}}+U\frac{\dy}{\dy{x}}\right)h+
 	H\left(\frac{\dy{u}}{\dy{x}}+\frac{\dy{v}}{\dy{y}}\right)+H'v & =0,
 	\label{ch3:linear-equ1e}
\end{align}
\end{subequations}
where a prime denotes differentiation. Eliminating for $v$, we obtain
\begin{equation}\label{ch3:SWMHDv}
	\left(\frac{S^{2}(Hv)'}{H(U-c)^{2}K^{2}}\right)' 
	-\left(\frac{\alpha^{2}S^{2}}{H(U-c)^{2}} -
	\frac{U'}{H(U-c)}\left(\frac{S^{2}}{(U-c)^{2}K^{2}}\right)' 
	+\frac{Q'S^{2}}{(U-c)^{3}K^{2}}\right)Hv=0,
\end{equation}
where $Q=-U'/H$ is the background potential vorticity, and 
\begin{equation}
   	S^{2}(y) = (U(y)-c)^{2} - M^2 B^{2}(y),\qquad K^{2}(y) = 1-F^{2}S^{2}(y). 
\end{equation}
Following \cite{Howard61}, under the transformation $Hv=(U-c)G$,
equation~\eqref{ch3:SWMHDv} becomes
\begin{equation}\label{ch3:SWMHD5}
	\left(\frac{S^{2}}{K^{2}}\frac{G'}{H}\right)'- \frac{\alpha^{2}S^{2}}{H} G=0. 
\end{equation}
We shall use this more compact form for the remainder of this study. In the
non-magnetic shallow-water limit ($M=0$), \eqref{ch3:SWMHDv} reduces to
equation~(3.4) of \cite{Balmforth99}. In the two-dimensional incompressible
magnetohydrodynamic limit ($F=0$ and $H=1$), \eqref{ch3:SWMHD5} reduces to
equation~(3.5) of \cite{HughesTobias01}.

We shall consider (\ref{ch3:SWMHD5}) in either an unbounded domain, for which
$|G| \rightarrow 0$ as $|y| \rightarrow \infty$, or in a bounded domain with
rigid side walls, where $G=0$ and hence $b_y=0$ via (\ref{ch3:linear-equ1d}).
Either way, for given real $\alpha$, (\ref{ch3:SWMHD5}) is then an eigenvalue
problem for the unknown phase speed $c=c_r+\zi c_i$. We will focus on
instabilities, i.e.\ $c_i \neq 0$, in which case (\ref{ch3:SWMHD5}) has no
singularities for real values of $y$. Since the transformation $\alpha
\rightarrow -\alpha$ leaves (\ref{ch3:SWMHD5}) unchanged, we may take $\alpha
\geq 0$ without loss of generality. Instability then occurs if $c_i > 0$, with
growth rate $\alpha c_i$.



\section{General theorems}

In this section we derive three results that hold for general shear flows
$U(y)$: two provide bounds on the growth rate of any instability, whereas the
third concerns implications of the parity of the basic state flow.


\subsection{Growth rate bound}

A bound on the instability growth rate may be obtained by calculating the rate
of change of the total disturbance energy using the combination
\begin{equation*}
	Hu^{*} \times \eqref{ch3:linear-equ1a} 
	+ Hv^{*} \times \eqref{ch3:linear-equ1b} 
	+ (M^2 Hb_x^{*}) \times	\eqref{ch3:linear-equ1c} 
	+ (M^2 H b_y^{*})\times\eqref{ch3:linear-equ1d} 
	+ h \times \eqref{ch3:linear-equ1e},
\end{equation*}
where $^{*}$ denotes complex conjugate. On adopting the form \eqref{ch3:expform}
for the perturbations, the real part of this expression gives (on dropping hats)
\begin{equation}\label{ch3:hoiland1}
\begin{aligned}
	&\alpha{}c_{i} \left( H \left(|u|^2 +|v|^2 +M^2 |b_x|^2 +M^2 |b_y|^2  \right) 
	+F^2 |h|^2 \right) = \\
	&-\Real\left( HU' \left(vu^{*} - M^2 b_x^{*} b_y \right)
	+ M^2 H B' \left( v b_x^{*} - u^{*} b_y \right) \right) -
	\Real\frac{\mathrm{d}}{\mathrm{d}y}(Hvh^{*}).
\end{aligned}
\end{equation}
On integrating over the $y$ domain, employing the boundary condition on $v$, and
manipulating the remaining terms on the right hand side using
$\pm2\Real(pq^{*})\leq|p|^{2}+|q|^2$, we obtain the following bound on the
growth rate:
\begin{equation}\label{thm:hoiland}
	\alpha{}c_{i}\leq\frac{1}{2}(\max|U'| + M \max|B'|).
\end{equation}
In the absence of magnetic field, this reduces to the well-known bound in
hydrodynamics \citep{Hoiland53,Howard61}.


\subsection{Semi-circle theorems}

In a classic paper, \cite{Howard61} proved that for incompressible hydrodynamic
parallel shear flows, the wave speed $c$ of any unstable mode must lie within a
semi-circle in the complex plane determined by properties of the basic state
flow. Subsequently, semi-circle theorems have been derived for several other
hydrodynamical and hydromagnetic systems \citep[e.g.,][]{CollingsGrimshaw80,
HayashiYoung87, ShivamoggiDebnath87, HughesTobias01}. In a similar manner, a
semi-circle theorem may be derived for the SWMHD system.

Multiplying equation~\eqref{ch3:SWMHD5} by $G^{*}$, integrating over $y$ and
using the boundary condition on $v$ (and hence $G$) gives the relation
\begin{equation}\label{ch3:SST1}
	\int \frac{S^2}{K^2} \frac{|G'|^2}{H} \, \mathrm{d}y 
	+ \alpha^2 \int \frac{S^2 |G|^2}{H} \, \mathrm{d}y = 0.
\end{equation}
The imaginary part of \eqref{ch3:SST1} gives
\begin{equation}\label{ch3:SWMHD6}
	c_{i}\int\ (U-c_{r})\chi \, \mathrm{d}y = 0,\qquad\textnormal{where}
	\qquad\chi=\frac{|G'|^{2}}{H|K|^{4}}+\alpha^{2} \frac{|G|^{2}}{H}\geq0.
\end{equation}
Equation~\eqref{ch3:SWMHD6} immediately yields Rayleigh's result that for
unstable modes ($c_i >0$), $c_r$ lies in the range of $U$ (i.e.\
$U_{\textnormal{min}} \leq {c_r} \leq U_{\textnormal{max}}$, where the
subscripts `min' and `max' refer to the minimum and maximum values across the
domain).

On using equation~\eqref{ch3:SWMHD6}, the real part of \eqref{ch3:SST1} gives
\begin{equation}\label{ch3:SWMHD7}
	(c_{r}^{2}+c_{i}^{2})\int\chi \, \mathrm{d}y 
	= \int\chi\left(U^{2} - M^{2} B^{2}\right) \, \mathrm{d}y
	-F^{2}\int\frac{|S|^{4}}{H|K|^{4}}|G'|^{2} \, \mathrm{d}y,
\end{equation}
which implies that
\begin{equation}\label{ch3:SWMHD8}
	0\leq(c_{r}^{2}+c_{i}^{2}) \int\chi \, \mathrm{d}y \leq 
	\left(U^{2} - M^{2}B^{2}\right)_{\textnormal{max}}
	\int\chi \, \mathrm{d}y.
\end{equation}
This gives the first semi-circle bound: the complex wave speed $c$ of an
unstable eigenfunction must lie within the region defined by
\begin{equation}\label{thm:semicircle1}
	c_r^2 + c_i^2 \leq\left(U^{2} - M^{2} B^{2}\right)_{\textnormal{max}}.
\end{equation}

The second semi-circle bound is obtained, in the standard manner, from the
inequality $0 \geq \int(U - U_{\textnormal{max}})(U - U_{\textnormal{min}}) \chi
\, \mathrm{d}y$. Substituting from \eqref{ch3:SWMHD6} and deriving an
inequality from \eqref{ch3:SWMHD7} leads to the expression
\begin{equation}\label{ch3:SWMHD10}
	0\geq \left( c_r^2 + c_i^2 - (U_{\textnormal{min}}+U_{\textnormal{max}})c_r 
	+ U_{\textnormal{min}}U_{\textnormal{max}} 
	+ M^2 (B^2)_{\textnormal{min}}\right) \int\chi\, \mathrm{d}y,
\end{equation}
which gives the second semi-circle bound: the speed $c$ of an unstable
eigenfunction must lie within the region defined by
\begin{equation}\label{thm:semicircle2}
	\left( c_{r} - \frac{U_{\textnormal{min}} 
	+ U_{\textnormal{max}}}{2} \right)^{2} +
	c_i^2  \leq\left( \frac{U_{\textnormal{max}} 
	- U_{\textnormal{min}}}{2} \right)^2
	- M^2 (B^2)_{\textnormal{min}}.
\end{equation}
Thus, taking these results together, the eigenvalue $c$ of an unstable mode must
lie within the intersection of the two semi-circles defined by
\eqref{thm:semicircle1} and \eqref{thm:semicircle2}. In the absence of magnetic
field, semi-circle~\eqref{thm:semicircle2} lies wholly within semi-circle
\eqref{thm:semicircle1}, and we recover the well-known result of
\cite{Howard61}. However, as observed by \cite{HughesTobias01}, who considered
the stability of aligned fields and flows in incompressible MHD, for non-zero
magnetic field there is the possibility of the two semi-circles overlapping,
being disjoint, or indeed ceasing to exist; thus, in addition to giving
eigenvalue bounds for unstable modes, these results also provide sufficient
conditions for stability. From \eqref{thm:semicircle1} and
\eqref{thm:semicircle2} it therefore follows that the basic state is linearly
stable if any one of the following three conditions is satisfied:
\begin{equation}\label{thm:lin-stable1}
	M|B|\geq|U|\qquad {\textnormal{everywhere in the domain}};
\end{equation}
\begin{equation}\label{thm:lin-stable2}
	M|B|_{\textnormal{min}}\geq
	 \frac{|U_{\textnormal{max}}-U_{\textnormal{min}}|}{2};
\end{equation}
\begin{equation}\label{thm:lin-stable3}
	\frac{U_{\textnormal{max}}+U_{\textnormal{min}}}{2} -
	\sqrt{\left(\frac{U_{\textnormal{max}}-U_{\textnormal{min}}}{2}\right)^{2} 
	+ M^2 (B^2)_{\textnormal{min}}} \geq
	\sqrt{\left(U^{2} - M^{2} B^{2}\right)_{\textnormal{max}}}.
\end{equation}
These results are equivalent to those given by \cite{HughesTobias01} for
incompressible MHD.

A drawback of the above approach is that the bounds do not contain the Froude
number $F$. Although it is possible to introduce $F$ into the semi-circle bounds
using similar manipulations to that employed by \cite{Pedlosky64a}, as shown by
\citet{Mak-thesis} this does not sharpen the bound and we thus omit it.


\subsection{Consequences of basic state parity}\label{parity}

For the hydrodynamic case, it can be shown that symmetries of the basic state
lead to symmetries in the stability problem \citep{Howard63}. These results may
be generalised to SWMHD if we make the further assumptions that $B^{2}(y)$ and
$H(y)$ are even functions about $y=0$.

We first consider the case when $U(y)$ is odd about $y=0$.
Equation~\eqref{ch3:SWMHD5} is unchanged under $c\rightarrow -c$ and $G(y)
\rightarrow G(-y)$. Since the equation is also unchanged under $c\rightarrow
c^*$ and $G \rightarrow G^*$, it follows that an eigenfunction with
eigenvalue $c= c_r + c_i$ must be accompanied by eigenfunctions with $c=\pm
c_{r} \pm \zi c_{i}$. Thus unstable solutions either have $c_r=0$ or are a pair
of counter-propagating waves with the same phase speed. As argued by
\cite{Howard63}, the symmetry in the basic state implies that there is no
preferred direction for wave propagation, consistent with the form of the
eigenvalues.

Now consider the case when $U(y)$ is even about $y=0$. Then
\begin{equation}
	G_{e} (y) = \frac{1}{2} \left( G(y)+G(-y) \right) 
	\qquad \textnormal{and} \qquad
	G_{o} (y) = \frac{1}{2} \left( G(y)-G(-y) \right)
\end{equation}
are also eigenfunctions of \eqref{ch3:SWMHD5}. Following \cite{DrazinHoward66},
if we now take $G_{o}$ multiplied by \eqref{ch3:SWMHD5} with $G=G_{e}$ and
subtract this from $G_{e}$ multiplied by \eqref{ch3:SWMHD5} with $G=G_{o}$,
integrating over $-L_y \le y \le L_y$ gives
\begin{equation}
W(G_e, G_o) \equiv [G_{e}'G_{o}-G_{o}'G_{e}]^{+L_{y}}_{-L_{y}} =
\textrm{constant} = 0,
\end{equation}
owing to the imposed boundary conditions on the eigenfunction. The vanishing of
the Wronskian $W$ implies that the functions $G_e$ and $G_o$ are linearly
dependent throughout the domain, which is possible only if one of them is
identically zero. Thus an unstable eigenfunction corresponding to a particular
eigenvalue is either an even or odd function about $y=0$.



\section{Piecewise-constant profiles: vortex sheet and rectangular jet}
\label{sec4}

We now consider some simple flow configurations for which the eigenvalue problem
\eqref{ch3:SWMHD5} can be reduced to an algebraic equation for $c$, from which
the conditions for stability can be readily determined. To do this, we take
$H=1$ (no topography) and $B=1$ (a uniform magnetic field). We seek solutions of
\eqref{ch3:SWMHD5} in an unbounded domain, with
\begin{equation}\label{ch4:bcs}
	|G| \rightarrow 0 \quad \textnormal{as} \quad |y| \rightarrow \infty.  
\end{equation}

We consider velocity profiles $U(y)$ that are piecewise constant. If $U(y)$ is
discontinuous at $y=y_0$, then the eigenfunction $G$ must satisfy two jump
conditions at $y=y_0$. In the usual way, the (linearised) kinematic boundary
condition implies 
\begin{subequations}
\begin{equation}\label{ch4:match1}
	\left[\frac{v}{U-c}\right]^{y_{0}^{+}}_{y_{0}^{-}} = 
	\left [ G \right]^{y_{0}^{+}}_{y_{0}^{-}}=0.
\end{equation}   
The pressure (or free surface displacement) is also continuous at $y=y_0$. The
corresponding condition on $G$ is most easily derived by integrating
\eqref{ch3:SWMHD5} across $y=y_0$, yielding
\begin{equation}\label{ch4:match2}
	\left[\frac{S^2}{K^2}G'\right]^{y_{0}^{+}}_{y_{0}^{-}}=0. 
\end{equation}
\end{subequations}


\subsection{Vortex sheet}\label{sec4_vs}

We first consider the velocity profile
\begin{equation*}
	U(y)=\begin{cases}+1, &y>0,\\-1, &y<0.\end{cases}
\end{equation*}
Then, for $y\neq0$, \eqref{ch3:SWMHD5} becomes $G''-\alpha^2 K^2 G = 0$. Using
\eqref{ch4:bcs} and \eqref{ch4:match1}, we thus find
\begin{equation}\label{ch4:vortex2}
	G(y)=\begin{cases} \exp \left( -\alpha{}K_{+}y \right), &y>0,\\
	                   \exp \left( +\alpha{}K_{-}y \right), &y<0,
	     \end{cases}
\end{equation}
where 
\begin{equation}\label{ch4:Kpm} 
 	K_{\pm}= \sqrt{ 1-F^2 \left( (1\mp{}c)^2-M^2 \right) }, 
 	\qquad \Real\left( K_{\pm} \right) > 0. 
\end{equation} 
The second jump condition \eqref{ch4:match2} then implies an eigenvalue relation
for $c$: 
\begin{equation}\label{ch4:vortex4}
	\frac{(1-c)^2-M^2}{K_{+}} + \frac{(1+c)^2-M^2}{K_{-}} = 0. 
\end{equation}
Note that $c$ is independent of the wavenumber $\alpha$, so any unstable mode
with $c_{i}>0$ has an unbounded growth rate as $\alpha \rightarrow \infty$. This
is an artefact of considering ideal fluids; viscosity will preferentially
suppress small scales and remove this unphysical behaviour.

There are several special cases. When $F=M=0$, we recover the classical
Kelvin--Helmholtz instability with $c=\pm\zi$. When $F=0$ but $M \neq 0$,
\eqref{ch4:vortex4} reduces to the incompressible MHD case of \cite{Michael55},
with $c^2=-(1-M^2)$; thus, the Kelvin--Helmholtz instability is stabilised when
$M \ge 1$, since the disturbance has to do work to bend the field lines. When
$M=0$ but $F \neq 0$, \eqref{ch4:vortex4} gives the classical hydrodynamic
shallow-water dispersion relation, which is analogous to that of two-dimensional
compressible hydrodynamics. The Kelvin--Helmholtz instability is stabilised when
$F\geq\sqrt{2}$ \citep{Miles58,BazdenkovPogutse83}, since the disturbance has to
do work to move the free surface against gravity. Thus, increasing $F$ or $M$ in
the absence of the other is stabilising. 

\begin{figure}
	\begin{center}
	\includegraphics[width=\textwidth]{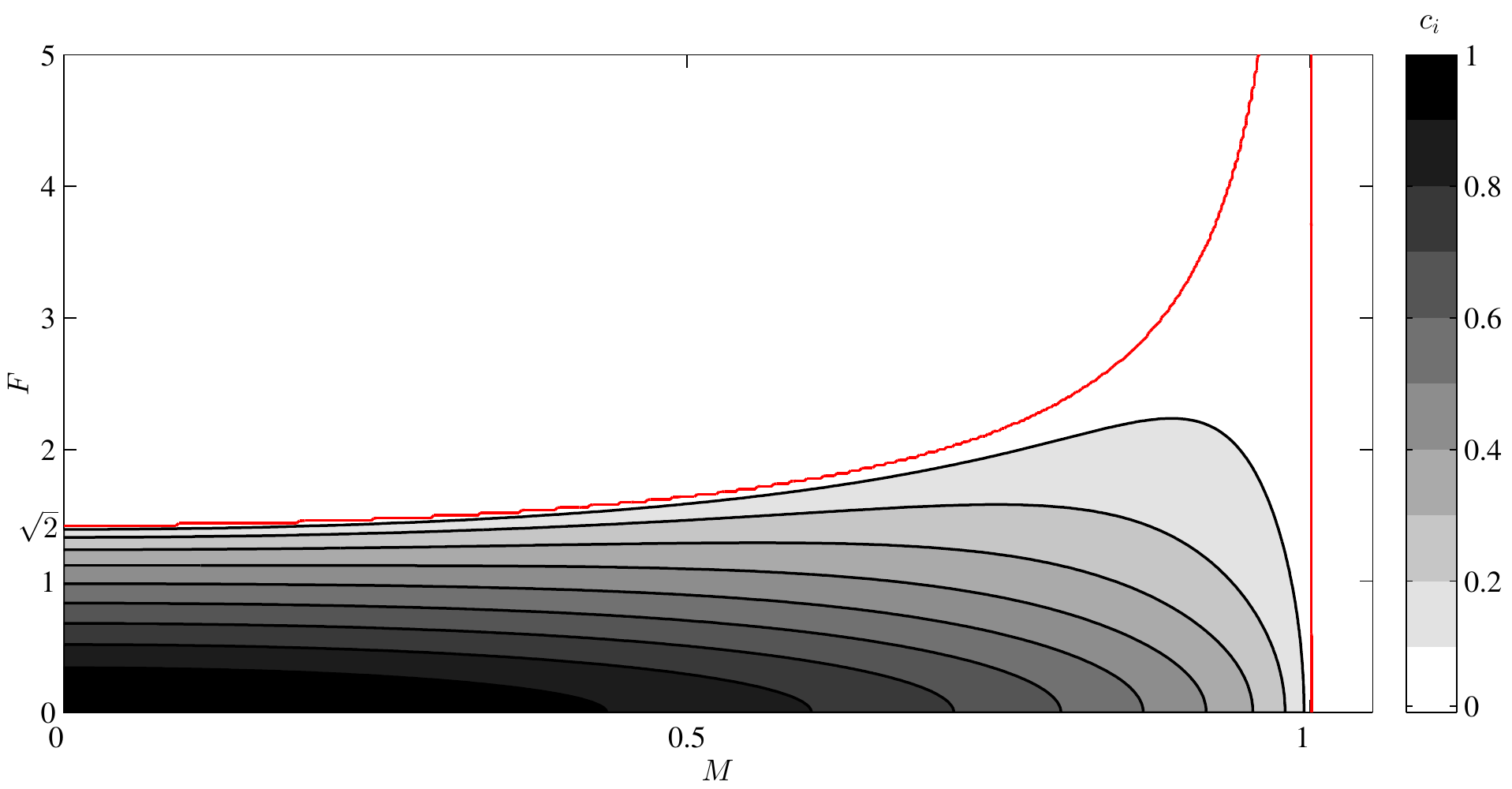}
	\caption{Contours of $\Imag(c_{\rm v})$, given by expression
	\eqref{ch4:vortex7}, with stability boundaries \eqref{ch4:vortex8} in red.}
	\label{fig:vortex_contour_FM}
	\end{center}
\end{figure}

In the general case where $F$ and $M$ are both non-zero, \eqref{ch4:vortex4} can
be rearranged and squared to yield a quartic equation for $c$: 
\begin{equation}\label{ch4:vortex5}
	F^2 c^4 - 2 \left( 1 + F^2 \left( M^2+1 \right) \right) c^2 
	+\left( M^2-1 \right) \left( 2 + F^2 \left( M^2-1 \right) \right) = 0. 
\end{equation}
Here we have ignored the degenerate case with $c=0$, which is a solution of
\eqref{ch4:vortex4} when $M=1$. By comparing solutions of \eqref{ch4:vortex5}
with those of \eqref{ch4:vortex4} found using a Newton iteration method, we find
that only two roots of \eqref{ch4:vortex5} also satisfy \eqref{ch4:vortex4}:
these are $c=\pm c_{\rm v}$, where
\begin{equation}\label{ch4:vortex7}
	c_{\rm v}=\zi\left( \frac{\sqrt{1+4F^2+4F^4M^2} 
	- (1+F^2+F^2M^2)}{F^2}\right)^{1/2}.
\end{equation}
A contour plot of $\Imag(c_{\rm v})$ is shown in
figure~\ref{fig:vortex_contour_FM}. From \eqref{ch4:vortex7}, there is
instability only if
\begin{equation}\label{ch4:vortex8}
	M<1 \qquad\textnormal{and}\qquad F<\sqrt{\frac{2}{1-M^2}}.
\end{equation}
Although increasing $F$ is always stabilising at fixed $M$, the critical value
of $F$ above which the flow is stable increases as $M$ increases towards $1$.
Thus, although magnetic field and free-surface effects are stabilising in
isolation, together they can lead to instabilities at arbitrarily large values
of $F$, provided
\begin{equation}\label{e:tongue}
  1 - \frac{2}{F^2} < M^2 < 1.
\end{equation}

\begin{figure}
  \begin{center}
  \includegraphics[width=\textwidth]{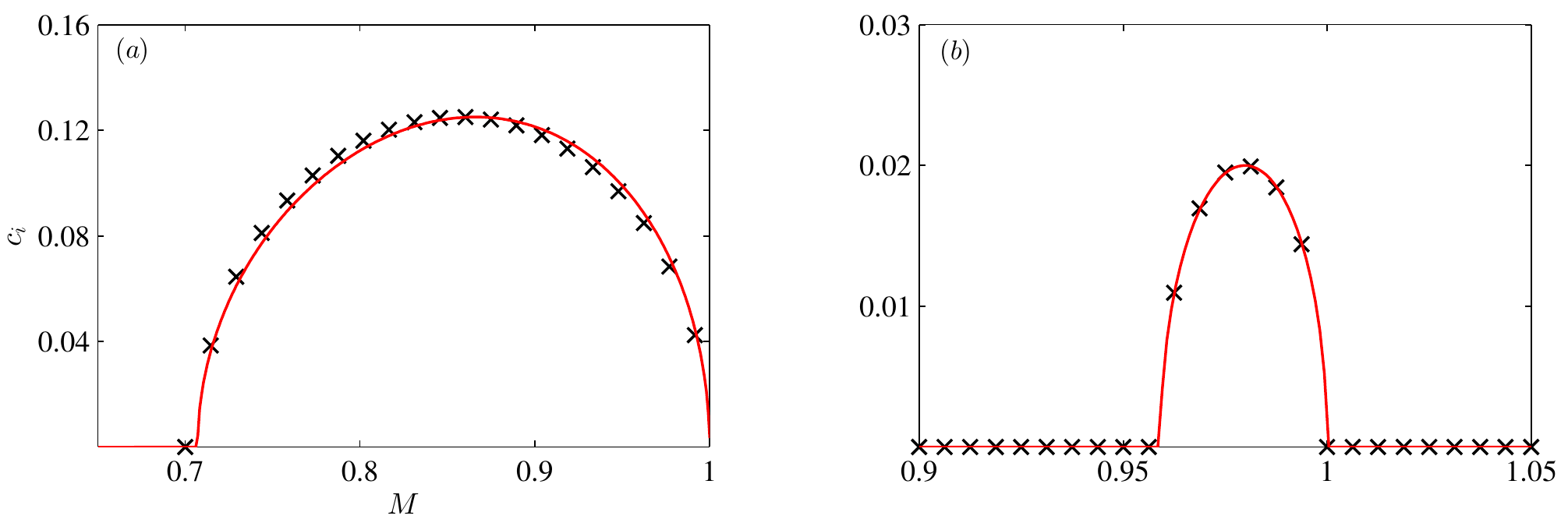}
  \caption{The weak instability regime of the vortex sheet for ($a$) $F=2$,
  ($b$) $F=5$, as determined directly from \eqref{ch4:vortex7} (crosses) and
  from the asymptotic result \eqref{ch4:vortex_asymp3} (line).}
  \label{fig:vortex_asymp}
  \end{center}
\end{figure}

Using an asymptotic analysis, it is possible to investigate these instabilities
further at large $F$ and with $M$ just smaller than unity. Rewriting
\eqref{ch4:vortex7} in terms of $1-M^2$ and expanding for $|1-M^2| \ll 1$, we
obtain
 \begin{equation}
  c_{\rm v} \sim \zi \left( \frac{1-M^2}{1+2F^2} 
  - \frac{2 F^6 (1-M^2)^2}{(1+2F^2)^3}\right)^{1/2}, \qquad |1-M^2| \ll 1, 
  \label{e:newcv}
 \end{equation}
where terms of $O \left( (1-M^2)^3 \right)$ have been neglected. When $F=O(1)$,
the first term on the right-hand side of \eqref{e:newcv} dominates. However, in
the regime of interest \eqref{e:tongue} with $F^2 \sim (1-M^2)^{-1} \gg 1$, the
two terms on the right-hand side of \eqref{e:newcv} have the same order of
magnitude, and instead we obtain
\begin{equation}\label{ch4:vortex_asymp3}
	c_{\rm v}\sim{}\zi\left( \frac{1-M^2}{2F^2}-\frac{(1-M^2)^2}{4} \right)^{1/2} 
	\quad\textnormal{as}\quad F^{-2}\sim(1-M^2)\rightarrow0.
\end{equation}
This simple formula is consistent with both stability boundaries in
\eqref{ch4:vortex8}, and, as shown in figure~\ref{fig:vortex_asymp}, closely
predicts $c_i$ in this weak instability regime, even when $F$ is of order unity.
Using \eqref{ch4:vortex_asymp3}, it is straightforward to show that
$\Imag(c_{\rm v})$ is maximised when $M^2 = 1 - 1/F^2$, with $c_{\rm v} \sim \zi
/ (2F^2)$, so that the growth rate of the most unstable mode decays like
$F^{-2}$ in this regime. 

\subsection{Rectangular jet}

We now consider the top-hat velocity profile
\begin{equation}\label{ch4:jet1}
	U(y)=\begin{cases}
		1, \ &|y|<1,\\
		0, \ &|y|>1.
	\end{cases}
\end{equation}
Then, \eqref{ch3:SWMHD5} and \eqref{ch4:bcs} imply
\begin{equation}\label{ch4:jet2}
	G=\left\{ \begin{array}{cc}
		A_{+}\exp \left( -\alpha{K_{0}}(y-1) \right), & \; \; y>+1,\\
		A_{e} \cosh(\alpha{K_{1}}y) +
		A_{o} \sinh(\alpha{K_{1}} y), & |y|<1,\\
		A_{-}\exp \left( +\alpha{K_{0}}(y+1) \right), & \; \; y<-1,
	\end{array} \right.
\end{equation}
for some $A_{+}$, $A_{-}$, $A_{e}$ and $A_{o}$, where 
\begin{subequations} \label{K0K1} \begin{align}
 	K_{0} & =\sqrt{ 1-F^2(c^2-M^2) },  \quad \textnormal{with} \; \; 
 	  \Real(K_0)>0 \; \; \mbox{for bounded solutions},\\  
	K_{1} & =\sqrt{1-F^2 \left( (1-c)^2-M^2 \right) }, \quad \mbox{with} \; \; 
	  -\frac{\pi}{2} < \arg \left( K_1 \right) \le \frac{\pi}{2}.
\end{align}\end{subequations}
Here we follow Rayleigh's formulation \citep{DrazinReid-Stability} and consider
eigenfunctions that are either even or odd. For the even mode, we set $A_{o}=0$,
$A_+ = A_-$ and write $c=c_e$. Then (\ref{ch4:match1},$b$) and \eqref{ch4:jet2}
give
\begin{equation}\label{ch4:jet_even_master}
  	\frac{c_e^2-M^2}{K_0} + \frac{(1-c_e)^2-M^2}{K_{1}}\tanh(\alpha K_{1} ) = 0.
\end{equation}
For the odd mode, we set $A_{e}=0$, $A_+ = -A_-$ and write $c=c_o$. Then
(\ref{ch4:match1},$b$) and \eqref{ch4:jet2} give
\begin{equation}\label{ch4:jet_odd_master}
  	\frac{c_o^2-M^2}{K_0} + \frac{(1-c_o)^2-M^2}{K_{1}}\coth(\alpha{}K_{1} ) = 0.
\end{equation}
In contrast to the vortex sheet dispersion relation \eqref{ch4:vortex4}, here
$c$ depends upon $\alpha$. 

Two special cases may be solved analytically. When $F=M=0$, so that $K_{0,1}=1$,
expressions~\eqref{ch4:jet_even_master} and \eqref{ch4:jet_odd_master} yield
unstable modes with
\begin{equation}
  c_{e} =\frac{T + \zi \sqrt{T}}{1+T}, \qquad
	c_{o} =\frac{1 + \zi \sqrt{T}}{1+T}, \qquad
	\mbox{where} \; \; T=\tanh \alpha, 
	\label{e:FM00}
 \end{equation}
so that the flow is unstable for all $\alpha$, with $\Imag(c)$ approaching a
maximum value of $1/2$ as $\alpha \rightarrow \infty$ \citep{Rayleigh78}. When
$F=0$ but $M \neq 0$, so that $K_{0,1}=1$ again, \eqref{ch4:jet_even_master} and
\eqref{ch4:jet_odd_master} yield
\begin{equation}\label{ch4:jet_F0_result}
    c_e=c_{e}^{(0)}=\frac{T + \zi \sqrt{T-M^2(1+T)^2}}{1+T}, \qquad
    c_o=c_{o}^{(0)}=\frac{1 + \zi \sqrt{T-M^2(1+T)^2}}{1+T}.
\end{equation}
(\citet{Gedzelman73} gave formulae for these two cases (his (3.4) and (3.5)),
but both are missing factors of $2$.) Both modes are stable for all $\alpha$
when $M \geq 1/2$; otherwise, both modes are unstable (with the same growth rate)
provided that
\begin{equation}
  M < M_c^{(0)} = \frac{\sqrt{T}}{1+T} \; \Leftrightarrow \;
 \alpha > \alpha_c^{(0)} 
 = \tanh^{-1} \left( \frac{1-2M^2-\sqrt{1-4M^2}}{2M^2} \right). 
 \label{e:Mc0alphac0}
\end{equation}
Thus, the magnetic field introduces a long-wave cutoff. 

The other special case with $M=0$ but $F \neq 0$ cannot be solved analytically.
Then, \eqref{ch4:jet_even_master} and \eqref{ch4:jet_odd_master} are equivalent
to expressions given by \cite{Gill65}, who considered corresponding
instabilities for two-dimensional compressible hydrodynamics. An important
property is that, at fixed $\alpha$ and $F$, there can be a large number of
unstable modes for sufficiently large $F$, which are often interpreted in terms
of over-reflection \citep{TakehiroHayashi92}. This is in contrast to
the behaviour at $F=0$, where just two modes exist (one even and one odd), as
described by \eqref{e:FM00}. 

\begin{figure}
\begin{center}
 \includegraphics[width=\textwidth]{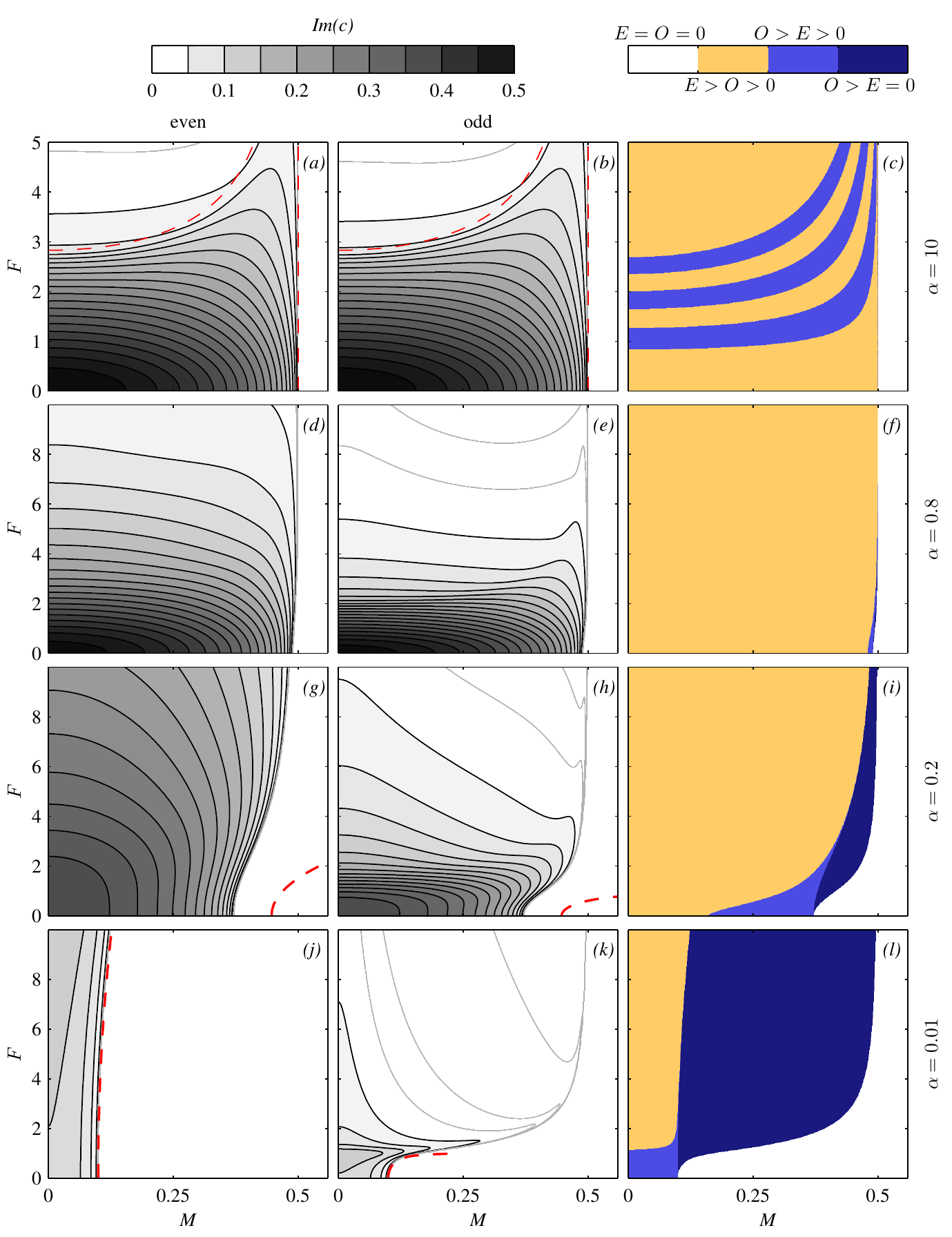}
	\caption{Instability of the rectangular jet. Smooth extensions at fixed
	$\alpha$ from $(M,F)=(0,0)$ of the unstable modes \eqref{e:FM00}, computed
	numerically using \eqref{ch4:jet_even_master} and
	\eqref{ch4:jet_odd_master}. Note the different choice of $F$-axis used in
	the top panels. Left: contours of $\Imag(c)$ for the even mode. Centre:
	contours of $\Imag(c)$ for the odd mode. The black contours are in intervals
	of $0.025$ from $0.025$ to $0.5$; weak instabilities are shown by the grey
	contours at $0$, $0.001$, $0.005$ and $0.01$. The $\alpha \gg 1$ stability
	boundaries \eqref{eq:conditions} are shown as dashed lines in
	panels ($a,b$). The $\alpha \ll 1$ stability boundaries
	\eqref{e:even_mode_cutoff2} and \eqref{e:odd_mode_cutoff} are shown as
	dashed lines in panels ($g,j$) and ($h,k$) respectively. Right: regime diagrams
	deduced from the left and centre panels, comparing the growth rate of the
	even mode ($E$) with that of the odd mode ($O$). No regions were found with
	$E > O = 0$. } 
	\label{fig:jet_contour_FM}
	\end{center}
\end{figure}

We generalise these results to cases with $F \neq 0$ and $M \neq 0$ by solving
\eqref{ch4:jet_even_master} and \eqref{ch4:jet_odd_master} numerically using a
Newton iteration. We start by limiting attention to the smooth extensions of the
unstable even and odd modes \eqref{e:FM00} that exist at $F=0$ and $M=0$, and
then tracking these modes over ($M,F$) space; the existence of other unstable
modes at high $F$ is discussed in \S\ref{s:highmodes}. Contours of $\Imag (c)$
obtained using this tracking approach are shown in
figure~\ref{fig:jet_contour_FM}, for a wide range of $F$, $M$ and $\alpha$. We
note that (i) neither mode is unstable for $M>1/2$; (ii) for small $\alpha$, the
even mode is stabilised at values of $M$ considerably less than $1/2$; (iii) for
small $\alpha$ and small $F$, the odd mode is stabilised at values of $M$
considerably less than $1/2$; (iv) both modes remain unstable for large $F$; (v)
when both modes are unstable, the even mode is generally more unstable than the
odd mode; (vi) at large $\alpha$, the even and odd modes lead to instabilities
of comparable strength, which mimic that of the vortex sheet (cf.
figure~\ref{fig:vortex_contour_FM}). We now use asymptotic analyses to describe
this behaviour in more detail. 

\subsubsection{Instability at large $\alpha$}\label{sec:large_alpha}

When $M$ and $F$ are of order unity, $\tanh(\alpha{}K_{1}) \approx 1$ when
$\alpha \gg 1$ (provided $\alpha \Real(K_1) \gg 1$), so that both
\eqref{ch4:jet_even_master} and \eqref{ch4:jet_odd_master} may be approximated
by
\begin{equation}\label{ch4:big_alpha_asymp1}
 \frac{ (\tilde c+1/2)^2-M^2 }{\sqrt{1-F^2((\tilde c+1/2)^2-M^2)}} 
 + \frac{(\tilde c-1/2)^2-M^2}{\sqrt{1-F^2\left( (\tilde c-1/2)^2-M^2 \right) }} 
 = 0, 
\end{equation}
where $\tilde c = c_e - 1/2$ or $\tilde c = c_o-1/2$. This dispersion relation
is similar in form to that for the vortex sheet \eqref{ch4:vortex4}, and a
solution may be found in the same way:
\begin{equation}\label{ch4:big_alpha_asymp2}
	\tilde{c} =\zi\left( \frac{\sqrt{1+F^2+F^4M^2 } -
	(1+F^2/4+F^2M^2)}{F^2} \right)^{1/2}.
\end{equation}
Physically, a sufficiently localised short-wave disturbance sees only one flank
of the jet, and thus a vortex sheet instability is obtained to a first
approximation. From \eqref{ch4:big_alpha_asymp2}, there is instability only when 
\begin{equation}\label{eq:conditions}
  	M<\frac{1}{2}\qquad\textnormal{and}\qquad F<\sqrt{\frac{8}{1-4M^2}}.
\end{equation}
As shown by the dashed lines in figures~\ref{fig:jet_contour_FM}$a,b$, the
conditions~\eqref{eq:conditions} approximately bound the region of strong
instabilities in ($M,F$) space, even when $\alpha = 10$. However, it is also
clear that there are additional weak instabilities (shown as grey contours) even
when $F > \sqrt{8/(1-4M^2)}$; the nature of these modes will be discussed in
\S\ref{s:highmodes}. In contrast, the $M=1/2$ cutoff is robust. Indeed, as is
evident from figures~\ref{fig:jet_contour_FM}$d,e$, the magnetic field cutoff
stays close to $M=1/2$ (for all $F$) even when $\alpha \approx 1$. For example,
at $\alpha=0.8$ and $F=0$, the cutoff $M_c \approx 0.490$, using
\eqref{e:Mc0alphac0}. 


\subsubsection{Instability at small $\alpha$}



We consider first the even mode. Assuming that $K_1$ remains bounded (which may
be confirmed \textit{a posteriori}), $\tanh(\alpha{}K_{1})\approx\alpha{}K_{1}$
when $\alpha \ll 1$, so that \eqref{ch4:jet_even_master} becomes
\begin{equation}\label{ch4:jet_even_asymp_master}
	\frac{c_e^2-M^2}{\sqrt{1-F^2(c_e^2-M^2)}} 
	+ \alpha \left( (1-c_e)^2-M^2 \right) = 0.
\end{equation}
Suppose that $F^2=O(1)$. If $M=O(1)$, then $c_e=\pm{}M$ at leading order, and
the next correction in $\alpha$ is also real: no instabilities are predicted,
consistent with figure~\ref{fig:jet_contour_FM}$j$. However, if $M^2 \sim
\alpha \ll 1$, then at leading order, \eqref{ch4:jet_even_asymp_master} gives
\begin{equation}\label{ch4:jet_even_asymp2}
	c_e \sim \pm \zi\sqrt{\alpha-M^2}\qquad\textnormal{as}\qquad 
	M^2\sim\alpha\rightarrow0,\quad F^2=O(1), 
\end{equation}
which is consistent with the small $M$ and $\alpha$ limits of
\eqref{ch4:jet_F0_result}. Thus a weak magnetic field reduces the strength of
the hydrodynamic instability and eventually suppresses it, with instability when
\begin{equation}\label{e:even_mode_cutoff1}
	M < M_c = \alpha^{1/2},\; \mbox{or equivalently when} \; 
   \alpha > \alpha_c = M^2.
\end{equation}
A dependence on $F$ appears only at the next order in $\alpha$, which is why the
cutoff in figure~\ref{fig:jet_contour_FM}$j$ is approximately independent of
$F$. 

\begin{figure}
	\begin{center}
	\includegraphics[width=\textwidth]{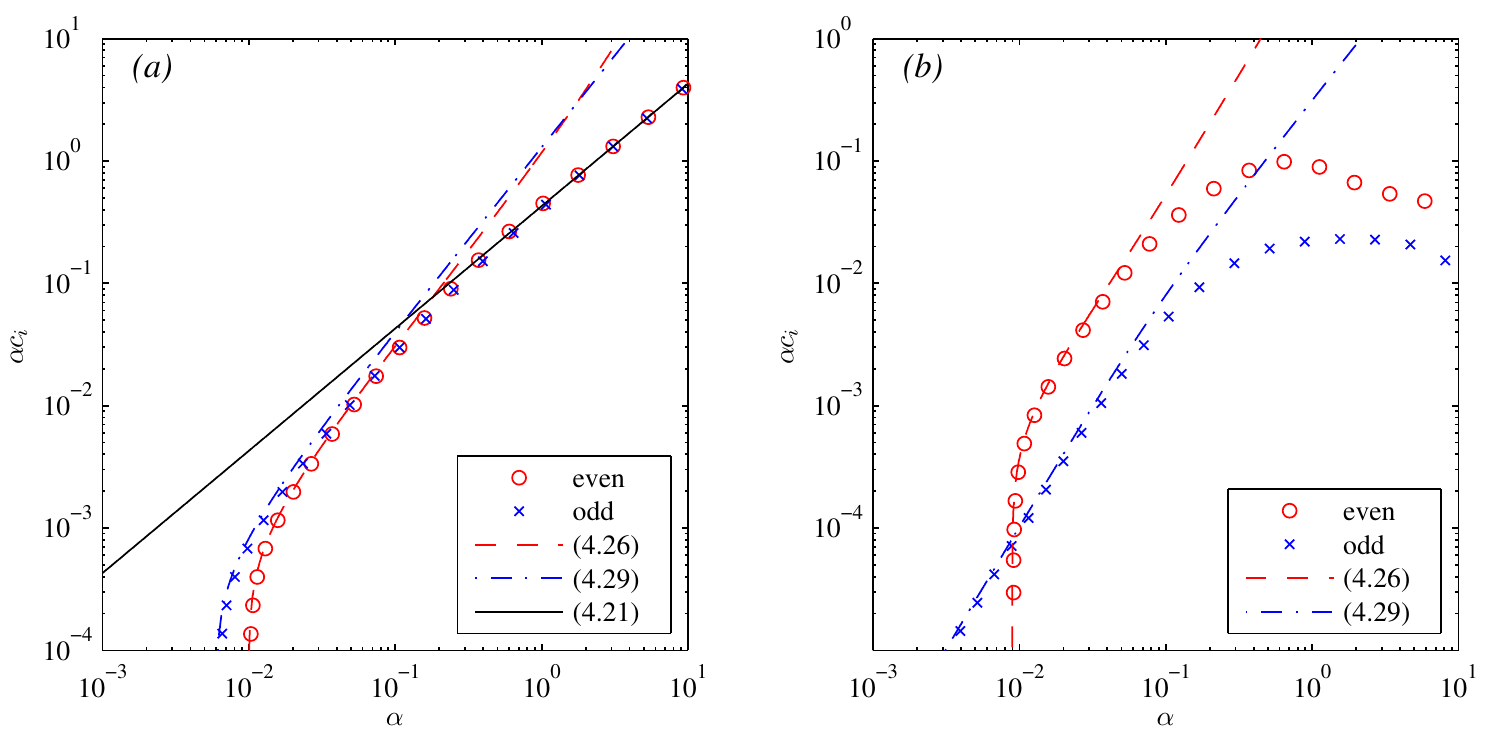}
	\caption{Instability of the rectangular jet at $M=0.1$, for ($a$) $F=0.8$
	and ($b$) $F=5$. Shown are the growth rates of the even and odd
	modes, computed numerically from \eqref{ch4:jet_even_master} and
	\eqref{ch4:jet_odd_master}, and the asymptotic predictions when $\alpha \ll
	1$ for the even mode \eqref{ch4:jet_even_asymp3} and the odd mode
	\eqref{ch4:jet_odd_asymp1}. In ($a$), the asymptotic prediction
	\eqref{ch4:big_alpha_asymp2} for both modes when $\alpha \gg 1$ is also
	plotted.}
	\label{fig:jet_alpha}
	\end{center}
\end{figure}

To capture the destabilising influence of $F$ apparent in
figure~\ref{fig:jet_contour_FM}$g$, the analysis may be extended to larger
values of $F$ with $F^2 \sim \alpha^{-1} \gg 1$. In this case, the square root
in \eqref{ch4:jet_even_asymp_master} also enters the balance at leading order,
and we obtain
\begin{equation}\label{ch4:jet_even_asymp3}
	c_e \sim\zi\left( \frac{\alpha^2F^2}{2}-M^2 
	+ \sqrt{\frac{\alpha^4F^4}{4}+\alpha^2}\right)^{1/2}
	\qquad\textnormal{as}\qquad M^2\sim{}F^{-2}\sim\alpha\rightarrow 0, 
\end{equation}
which formally reduces to \eqref{ch4:jet_even_asymp2} when $F$ is of order
unity. Again there is a long-wave cutoff due to the magnetic field, with
instability only when
 \begin{equation}
   M < M_c = \frac{ \alpha^2 F^2 + \sqrt{4 \alpha^2 + \alpha^4 F^4} }{2} 
   \; \; \Leftrightarrow \; \; 
   \alpha > \alpha_c= \frac{M^2}{\sqrt{1+M^2 F^2}}. 
  \label{e:even_mode_cutoff2}
 \end{equation}  
This cutoff, which is shown as a dashed contour in
figures~\ref{fig:jet_contour_FM}$g,j$, captures the destabilising influence of
$F$. Indeed, in contrast to the result \eqref{eq:conditions} for $\alpha \gg 1$,
equation \eqref{ch4:jet_even_asymp3} does not predict stabilisation at large $F$
when $\alpha \ll 1$, consistent with the results shown in
figures~\ref{fig:jet_contour_FM}$g,j$. The accuracy of this dispersion
relation and cutoff when $\alpha \ll 1$ can be seen in
figure~\ref{fig:jet_alpha}. 

In the same $\alpha \ll 1$ limit, the dispersion relation
\eqref{ch4:jet_odd_master} for the odd mode reduces to
\begin{equation}\label{ch4:jet_odd_asymp_master}
	\frac{\alpha(c_o^2-M^2)}{\sqrt{1-F^2(c_o^2-M^2)}} 
	+ \frac{(1-c_o)^2-M^2}{1-F^2 \left( (1-c_o)^2-M^2 \right) }=0. 
\end{equation}
Suppose that $F^2=O(1)$. If $M=O(1)$, then $c_o$ is real at the first two orders
in $\alpha$: no instabilities are predicted, consistent with
figure~\ref{fig:jet_contour_FM}$k$. However, if $M^2 \sim \alpha \ll 1$, then
at leading order, \eqref{ch4:jet_odd_asymp_master} gives 
\begin{equation}\label{ch4:jet_odd_asymp1}
	c_o \sim 1+\zi \left( \frac{\alpha}{\sqrt{1-F^2}} - M^2 \right)^{1/2} 
        \; \textnormal{as} \; \; 
	M^2\sim\alpha\rightarrow0,\; F^2=O(1),
	\; 1-F^2=O(1).  
\end{equation}
When $F<1$, again there is a cutoff due to the magnetic field, with instability
only when
 \begin{equation}
   M < M_c = \frac{ \alpha^{1/2} }{ (1-F^2)^{1/4}}, 
   \; \; \mbox{or equivalently when} \; \;
   \alpha > \alpha_c = M^2 \sqrt{1-F^2}. 
   \label{e:odd_mode_cutoff}
  \end{equation} 
This cutoff, which is shown as a dashed contour in
figures~\ref{fig:jet_contour_FM}$h,k$, captures the sharp destabilising
transition as $F \rightarrow 1$, which is particularly evident in
figure~\ref{fig:jet_contour_FM}$k$. However, when $F>1$,
expression~\eqref{ch4:jet_odd_asymp1} shows that $c_o$ always has a positive
imaginary part, so that there is no cutoff at small $M$ (when $\alpha \ll 1$).
This is clear in figures~\ref{fig:jet_contour_FM}$h,k$, where there are weak
instabilities for large $F$ with $M$ close to $1/2$. The absence of a cutoff
when $F>1$ is also evident in figure~\ref{fig:jet_alpha}$b$. 

When $F$ is close to $1$, the asymptotics leading to \eqref{ch4:jet_odd_asymp1}
break down, and one must instead seek solutions with $c = 1 + O(\alpha^{2/5})$.
The resulting (quintic) equation does not admit solutions in closed form
\citep{Mak-thesis}, although the special case with $F=1$ may be solved exactly,
yielding $c_o = 1+\alpha^{2/5} {\rm e}^{3 \pi \zi / 5} / 2^{1/5}$. Thus $c_o$ is
independent of $M$ in this regime, consistent with the behaviour shown in the
bottom left corners of figures~\ref{fig:jet_contour_FM}$h,k$. 

\subsubsection{Preferred mode of instability: even versus odd modes}

From figure~\ref{fig:jet_contour_FM}, it can be seen that the even mode is more
unstable than the odd mode for some parameters but not for others. When $\alpha
\gg 1$, both even and odd modes satisfy the same dispersion relation
\eqref{ch4:big_alpha_asymp1} to leading order, as seen in
figures~\ref{fig:jet_contour_FM}$a,b$; as shown in
figure~\ref{fig:jet_contour_FM}$c$, the even mode is generally more unstable
than the odd mode, although this is a weak effect. In contrast, when $\alpha \ll
1$, the odd mode is generally more unstable, as shown in
figure~\ref{fig:jet_contour_FM}$l$. In particular, when $F<1$, the odd mode is
more unstable (from \eqref{ch4:jet_even_asymp2} and \eqref{ch4:jet_odd_asymp1})
and is unstable over a larger region of parameter space (from
\eqref{e:even_mode_cutoff1} and \eqref{e:odd_mode_cutoff}). When $F>1$, again
the odd mode is unstable over a larger region of ($M,F$) space; the even mode is
stabilised according to \eqref{e:even_mode_cutoff2}, whilst the odd mode has no
cutoff at small $M$, from \eqref{e:odd_mode_cutoff}. 

The nature of the transition when $\alpha$ is of order unity can be quantified
by performing a small $F$ analysis. We thus write $c=c^{(0)}+F^2c^{(1)}+O(F^4)$
and substitute into \eqref{ch4:jet_even_master} for $c_e$ and
\eqref{ch4:jet_odd_master} for $c_o$. The leading order terms $c_e^{(0)}$ and
$c_o^{(0)}$ are given by \eqref{ch4:jet_F0_result}. In the case where there is
instability at leading order, i.e.\ when \eqref{e:Mc0alphac0} is satisfied, we
then find  
 \begin{align}
 \begin{split}
   &\Imag \left( c_e^{(1)} \right) = a_e g, \; \; \;
   \Imag \left( c_o^{(1)} \right) = a_o g, \; \; \;
   \mbox{where} \;  \;  g(\alpha,M) =  \frac{T^2 - 6T + 1}{(T+1)^2} + 4M^2, \\
   &a_e = \frac{T+T^2 -\alpha(1-T^2)}{4 (1+T)^2 \sqrt{T-M^2(1+T)^2}}, \; \; \; 
   a_o = \frac{T+T^2 + \alpha T (1-T^2)}{4 (1+T)^2 \sqrt{T-M^2(1+T)^2}}, 
 \end{split}
 \end{align} 
and $T=\tanh \alpha$. It is easy to see that $a_o > a_e > 0$ for all $\alpha$,
so the behaviour is determined by the sign of $g$. More precisely, the even mode
is more unstable (at small $F$) when $4M^2 < (6T-1-T^2)/(T+1)^2$. This is
impossible when $T < 3-2\sqrt{2}$, so the odd mode is more unstable at small $F$
when $\alpha < 0.1733$, for all $M$. If $T > 3-2 \sqrt{2}$, then the even mode
is more unstable at small $F$ when
 \begin{equation}
  M < M_c = \frac{ \sqrt{ 6T-1-T^2} }{2 (1+T) } 
  \; \Leftrightarrow \; 
  \alpha < \tanh^{-1}\left(\frac{3-4M^2 - 2\sqrt{2-4M^2}}{1+4M^2}\right).
 \end{equation} 
For example, when $\alpha=0.2$, the most unstable mode at small $F$ is even when
$M < 0.159$ and odd when $0.159 < M < 0.371$ (cf.
figure~\ref{fig:jet_contour_FM}$i$); when $\alpha=0.8$, the most unstable mode
is even when $M < 0.479 $ and odd when $0.479 < M < 0.490 $ (cf.
figure~\ref{fig:jet_contour_FM}$f$). As $\alpha \rightarrow \infty$, $T
\rightarrow 1$ and $M_c \rightarrow 1/2$, so there is only a vanishingly thin
region close to $M=1/2$ where the odd mode is more unstable at small $F$. 


\subsubsection{Maximum growth rate}

At fixed $F$ and $M$, it is natural to try to find the most unstable wavenumber
by maximising $\alpha c_i$ with respect to $\alpha$. However, this is not
guaranteed to lead to a finite value of $\alpha$: for example, when $F=0$,
\eqref{ch4:jet_F0_result} shows that $c_i \rightarrow \sqrt{1-4M^2}/2$ as
$\alpha \rightarrow \infty$, so provided $M<1/2$ the growth rate $\alpha c_i$
increases without bound as $\alpha \rightarrow \infty$. The same phenomenon
persists for small $F$, as illustrated in figure~\ref{fig:jet_alpha}$a$ at
$M=0.1$ and $F=0.8$. However, when $F > \sqrt{8/(1-4M^2)}$ and $M<1/2$, the
asymptotic theory of \S\ref{sec:large_alpha} for $\alpha \gg 1$ predicts that
the flow is stable. In practice, we have seen in
figures~\ref{fig:jet_contour_FM}$a,b$ that there are weak instabilities for
these parameters, but the growth rates decay sufficiently rapidly with $\alpha$
that a most unstable mode typically occurs at finite $\alpha$, as illustrated in
figure~\ref{fig:jet_alpha}$b$ at $M=0.1$ and $F=5$. However, we have been unable
to find an analytical expression for the most unstable wavenumber in this
regime. 


\subsubsection{Multiple modes of instability at large $F$}\label{s:highmodes}

\begin{figure}
	\begin{center}
	\includegraphics[width=\textwidth]{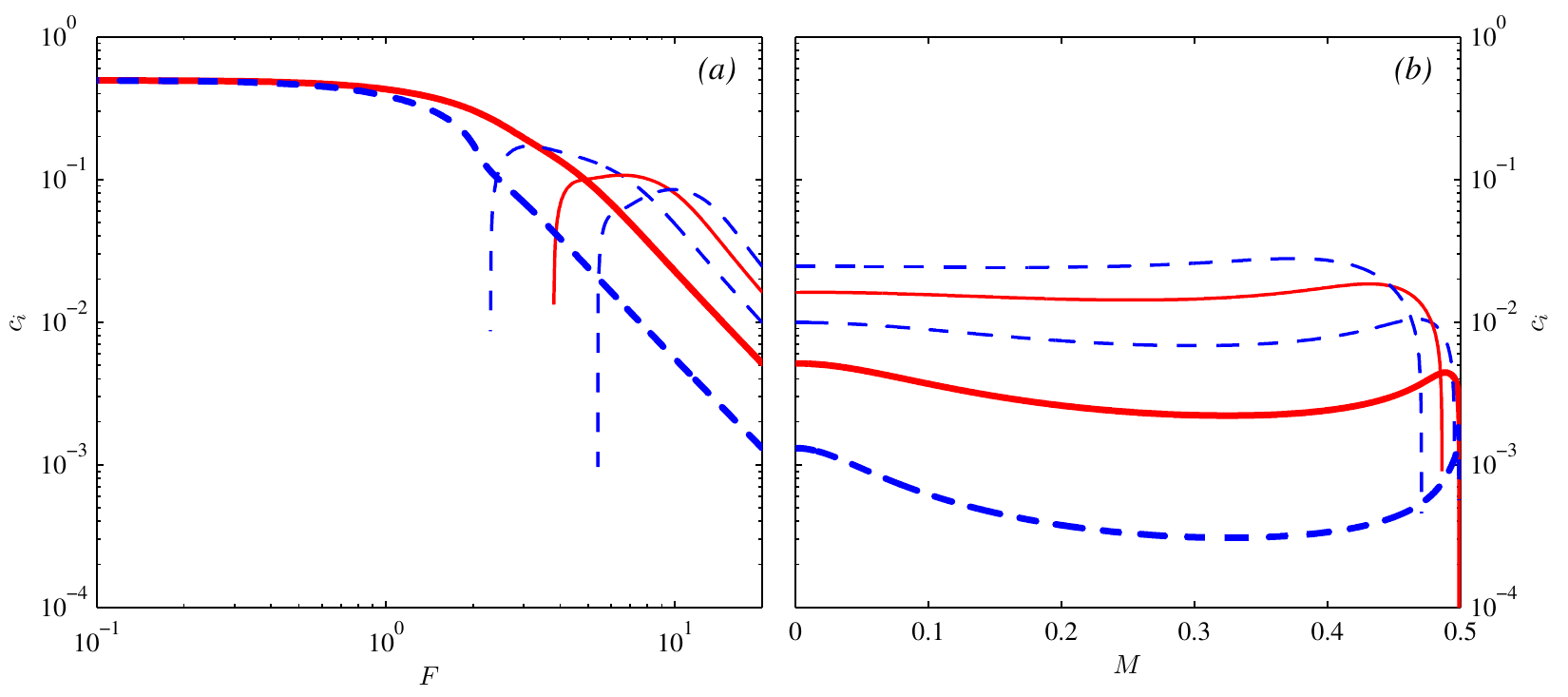}
	\caption{Instability of the rectangular jet at $\alpha=1$, and $M=0$ (left)
	and $F=20$ (right). Shown is $c_i=\Imag(c)$ computed numerically from
	\eqref{ch4:jet_even_master} and \eqref{ch4:jet_odd_master}. Even modes are
	shown as solid lines, and odd modes as dashed lines. The unique even and odd
	modes that smoothly connect to the modes at $(M,F)=(0,0)$ are shown as
	thicker lines.}
	\label{fig:gill}
	\end{center}
\end{figure}

In addition to the smooth extensions of the unstable even mode and odd mode that
exist at $(M,F)=(0,0)$ (the primary modes), there exist secondary modes of
instability for sufficiently large $F$. Instabilities of this type are shown in
figure~\ref{fig:gill}. At $M=0$, the first such secondary mode, which is odd,
appears close to $F=2.3$; the next secondary mode, which is even, appears close
to $F=3.8$. For larger values of $F$, the secondary modes are more unstable than
the primary modes; this behaviour persists for $M \neq 0$, as shown in
figure~\ref{fig:gill}$b$. Although an asymptotic description of the secondary
modes is possible as $F \rightarrow \infty$, the growth rates are relatively
small, so these modes are not discussed further. 



\section{Smooth profiles: hyperbolic-tangent shear layer and Bickley jet}
\label{sec:smooth_profiles}

As a model of more realistic velocity profiles, we consider in this section the
instabilities of the hyperbolic-tangent shear layer and the Bickley jet. These
may be regarded as smooth versions of the piecewise-linear profiles studied in
\S\ref{sec4}. Linear instability calculations involving these two profiles are
well documented in a wide variety of contexts \citep[e.g.,][]{Lipps62, Howard63,
Michalke64, SutherlandPeltier92, HughesTobias01} and these provide a comparison
and check on our results. We again restrict attention (for simplicity) to the
case of a uniform background magnetic field, $B(y) \equiv 1$, and with no
underlying topography.


\subsection{Numerical method}

We seek a numerical solution of the eigenvalue equation \eqref{ch3:SWMHD5},
written as
\begin{equation}\label{ch5:numerical_1}
	G''+\left( \frac{(S^2)'}{S^2} - \frac{(K^2)'}{K^2} \right) G'
	-\alpha^2 K^2 G=0,
\end{equation}
where, again, $S^2=(U-c)^2 - M^2$ and $K^2 = 1-F^2 S^2$. Although the velocity
profiles are defined over the entire real line, we solve \eqref{ch5:numerical_1}
on $y\in [-L, L]$, using a shooting method, with matching imposed at $y=0$,
employing a generalised Newton method as the root-finding algorithm. Since the
solutions decay exponentially as $|y|$ becomes large, namely
\begin{equation}\label{ch5:numerical_2}
	G\sim\exp \left(-\alpha{}K_{\pm} y \right) 
	\qquad \textnormal{as} \qquad |y|\rightarrow \infty,
\end{equation}
where
\begin{equation}
	K_{\pm}^2 = 1-F^2 \left( (U_{\pm} - c)^2 - M^2 \right) , 
	\qquad U_{\pm} = \lim_{y\rightarrow\pm\infty} U(y), 
\end{equation}
we adopt expression~\eqref{ch5:numerical_2} as the boundary condition to be
implemented at $y=\pm L$. Since our interest is in instabilities, singularities
in the governing equation are avoided. To ensure negligible influence of the
finite domain, $L$ is doubled until the computed eigenvalue changes by less than
0.5\%. The routines are written in MATLAB, using \verb|ode113| as the integrator
(an Adams-Bashforth method with adaptive grid). Although the boundary conditions
are functions of $c$, changing at every iteration, we generally have no problems
with convergence provided that the initial guess is close to the true value.
Solutions are initialised from $(M,F)=(0,0)$ at some fixed $\alpha$ using a
known numerical result documented in, for example, \cite{DrazinReid-Stability}.
Runs at new parameter values are then initialised using an estimate for $c$ from
previously calculated values at nearby parameters. The Bickley jet is even about
$y=0$ and hence the parity result of \S\ref{parity} holds --- i.e. the
eigenfunctions are either even or odd. In this case we need integrate only up to
$y=0$, with the imposition of either $G'(0)=0$ (even mode) or $G(0)=0$ (odd
mode).


\subsection{Hyperbolic-tangent shear layer}

In this subsection, we  consider the basic state velocity defined by
\begin{equation}
	U(y)=\tanh y.
  \label{e:tanh}
\end{equation}
From inequality~\eqref{thm:hoiland} we know that the growth rates
$\alpha{}c_{i}$ are bounded above by $|U'|_{\mathrm{max}}/2=1/2$; furthermore,
from the stability criteria \eqref{thm:lin-stable1} or \eqref{thm:lin-stable2},
this profile is stable when $M\geq1$.

When $M=F=0$, instability exists only when $0 < \alpha < 1$, with a neutral mode
at $\alpha=1$ \citep[e.g.,][\S31.10]{DrazinReid-Stability}. in this case, there
is a single primary mode of instability, which may be classified as an
inflection-point instability and attributed to interacting Rossby waves
supported by the background shear \citep[see, for example, the review
by][]{Carpenter-et-al13}. For both two-dimensional compressible hydrodynamics
and shallow-water hydrodynamics, there is a secondary mode of instability, first
found by \cite{Blumen-et-al75}. This has a smaller growth rate and a less
pronounced spatial decay than the primary mode. Further, whilst the primary mode
has $c_r=0$, there are two branches for the secondary mode with equal and
opposite (non-zero) phase speeds, consistent with the parity results in
\S\ref{parity}, which ensure that $c=\pm c_r+\zi c_i$ for unstable modes. The
secondary mode can be attributed to interacting gravity waves
\citep[e.g.,][]{Satomura81, HayashiYoung87, TakehiroHayashi92, Balmforth99} and,
indeed, can occur for linear shear flows, explicitly filtering out the
possibility of Rossby waves due to a background vorticity gradient. This is
consistent with the theorem of \cite{Ripa83}, which states that for instability,
either the associated potential vorticity profile possesses an inflection point
or $F>1$, both of which encourage instability through interacting waves.

\begin{figure}
	\begin{center}
	\includegraphics[width=\textwidth]{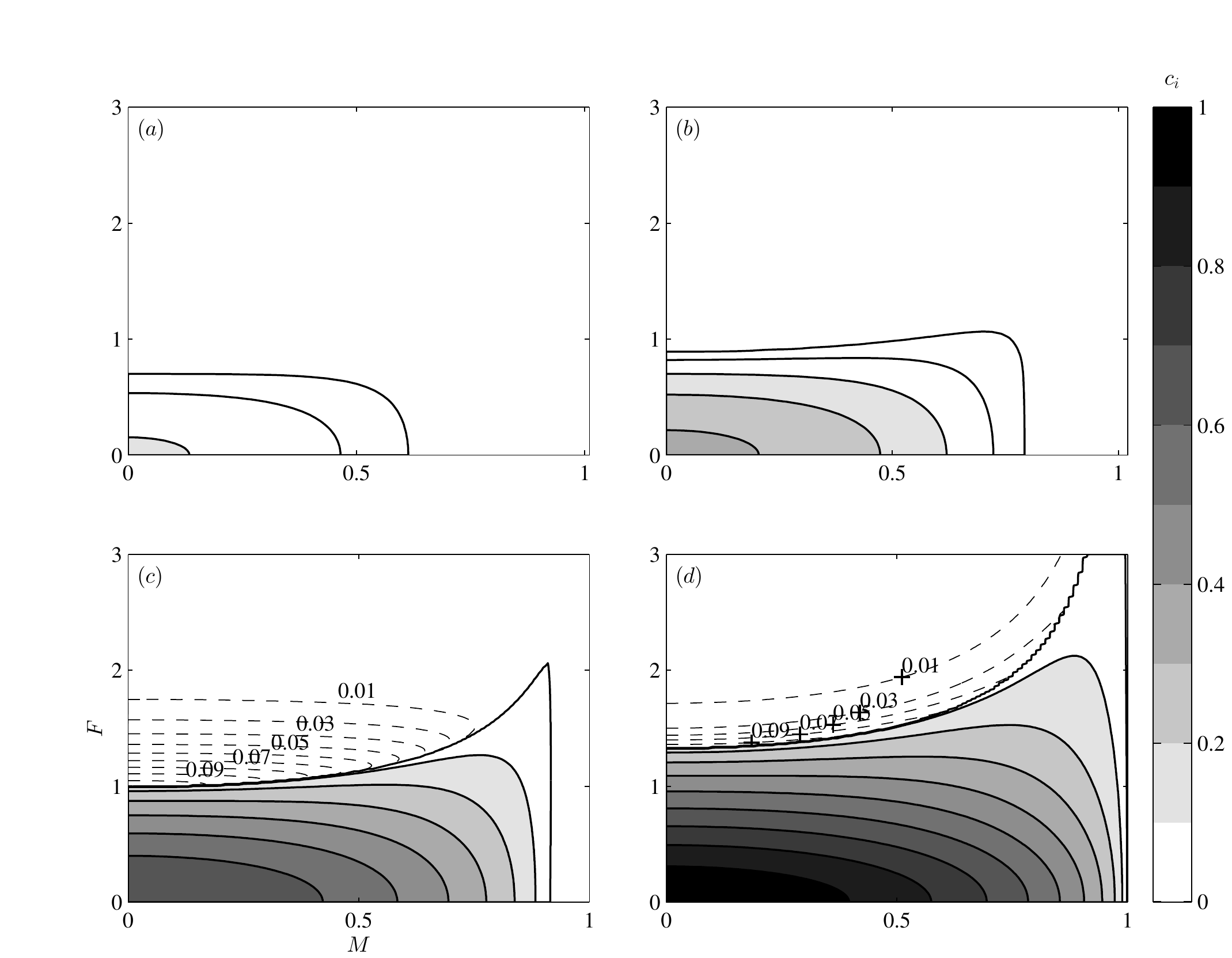}
	\caption{Contours of $c_i$ over $(M, F)$ space at ($a$) $\alpha=0.70$, ($b$)
	$\alpha=0.44$, ($c$) $\alpha=0.20$, ($d$) $\alpha=0.01$. The primary modes
	(here classified by $|c_r| < 10^{-3}$) are contoured as solid lines. The
	secondary modes are contoured as dashed lines.}
	\label{fig:tanh_contour_FM}
	\end{center}
\end{figure}

Both modes have been found in our SWMHD system. Figure~\ref{fig:tanh_contour_FM}
shows contours of $c_i$ over $(M,F)$ space at selected values of $\alpha$ for
the profile (\ref{e:tanh}), distinguishing between the primary mode (with
$c_r=0$, shown as solid contours) and the secondary mode (with $c_r \neq 0$,
shown as dashed contours). In figure~\ref{fig:tanh_contour_FM}$a$, at
$\alpha=0.7$, and figure~\ref{fig:tanh_contour_FM}$b$, at $\alpha=0.44$ (which
is the most unstable mode when $M=F=0$ \citep{Michalke64}), only the primary
mode exists. In figure~\ref{fig:tanh_contour_FM}$c$, at $\alpha=0.2$, and
figure~\ref{fig:tanh_contour_FM}$d$, at $\alpha=0.01$, both modes exist,
although largely in different parts of $(M,F)$ space. Note that
figure~\ref{fig:tanh_contour_FM}$d$ and figure~\ref{fig:vortex_contour_FM} are
remarkably similar, suggesting that long-wave instabilities for this velocity
profile resemble those of the vortex sheet. This will be quantified via a
long-wave asymptotic analysis in \S\ref{sec:LWA}, in which we derive the more
general result that long-wave instabilities of \textit{any} shear-layer profile
display the characteristics of vortex sheet instabilities.

\begin{figure}
	\begin{center}
	\includegraphics[width=\textwidth]{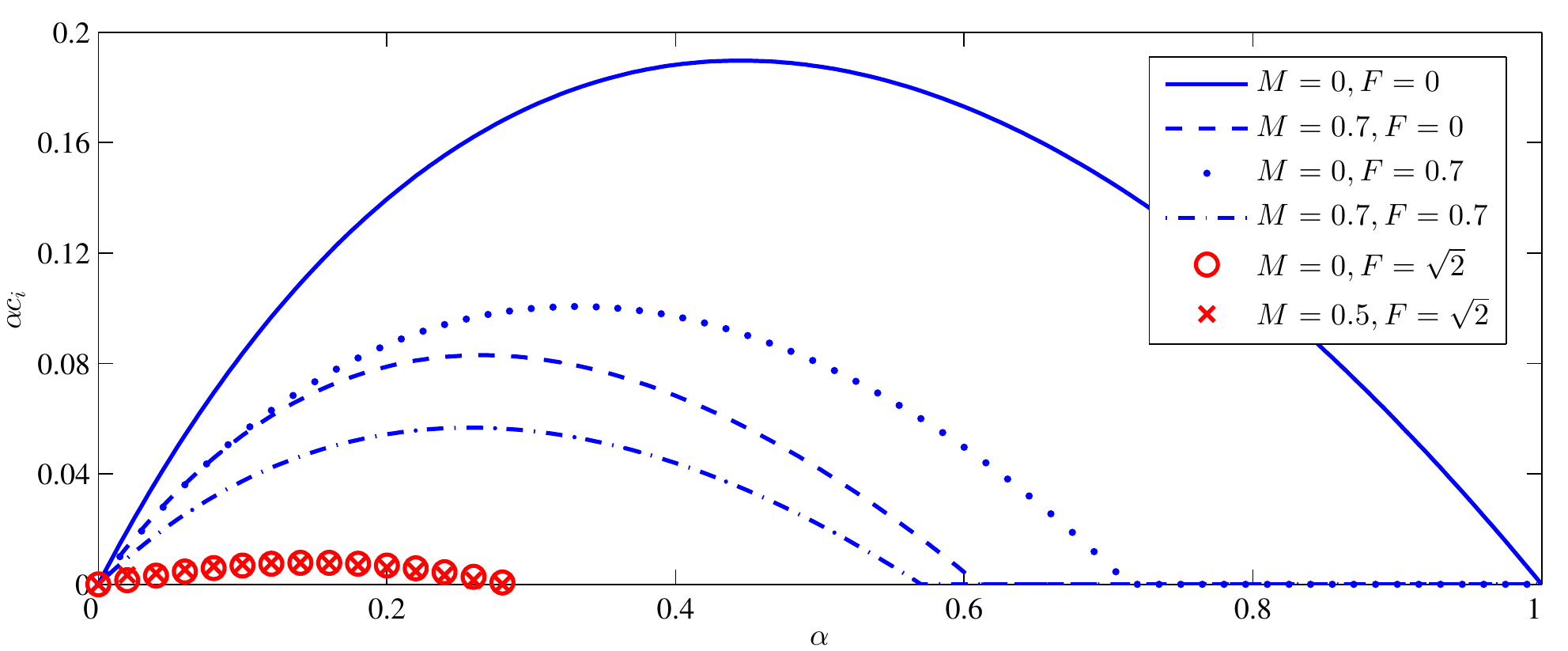}
	\caption{Growth rate versus $\alpha$ at selected parameter values for
	$U(y)=\tanh y$. The upper four curves are primary modes, the lower two are
	secondary modes.}
	\label{fig:tanh_growth}
	\end{center}
\end{figure}

The growth rate $\alpha c_i$ is shown in figure~\ref{fig:tanh_growth} as a
function of $\alpha$. It can be seen that the secondary modes generally have
weaker growth rates than the primary modes, consistent with the results of
\cite{Blumen-et-al75}. As we shall see in \S\ref{sec:LWA}, the relation between
the two types of unstable modes can be explored in some detail in the long
wavelength limit.

\subsection{Instability mechanism: Counter-propagating Rossby Waves}

As mentioned above, inflection point instabilities can be attributed to
interacting Rossby waves supported by the background shear. The constructive
interference of a pair of Counter-propagating Rossby Waves (CRWs) has been
proposed as the mechanism leading to instability of shear flows in a variety of
settings \citep[e.g.,][]{Bretherton66a, Hoskins-et-al85, BainesMitsudera94,
Heifetz-et-al04a, HarnikHeifetz07}. For the SWMHD system, it is therefore
natural to enquire how this underlying mechanism is modified by magnetic and
shallow-water effects.

\begin{figure}
	\begin{center}
	\includegraphics[width=\textwidth]{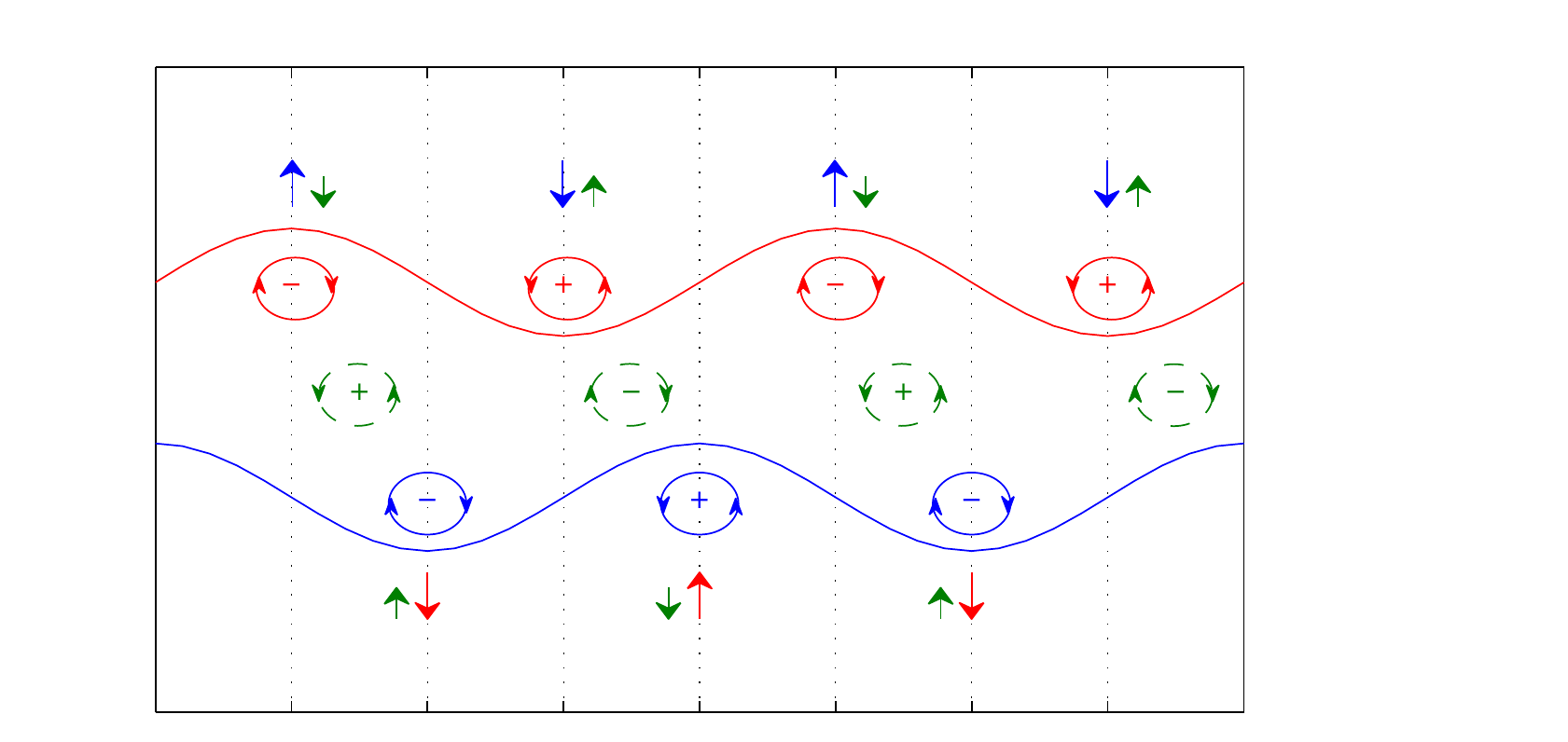}
	\caption{Modified CRW mechanism in schematic form. Shown are two displaced
	material contours. The associated vorticity anomalies when $M=F=0$ are shown
	by the closed solid curves; the effect of these on the other contours, leading
	to instability, is shown by the longer arrows. The closed dashed curves
	represent the additional vorticity anomalies when $M$ and $F$ are non-zero.
	The stabilising effect of these, which opposes the driving of the instability,
	is shown by the shorter arrows.}
	\label{fig:CRWmod}
	\end{center}
\end{figure}

\begin{figure}
	\begin{center}
	\includegraphics[width=\textwidth]{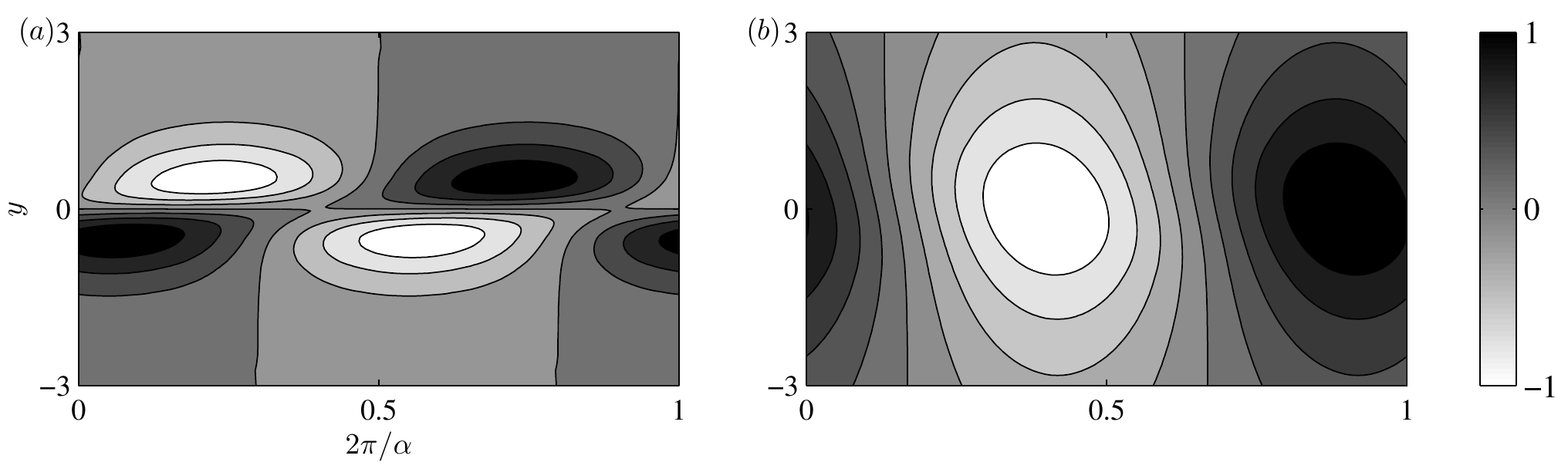}
	\caption{Eigenfunctions of ($a$) vorticity and ($b$) height for the most
	unstable modes for $U(y)=\tanh y$ at $F=0$, $M=0$.}
	\label{fig:tanh_most_unstable_eigenfunction}
	\end{center}
\end{figure}

Let us first consider the case of $M = 0$, $F = 0$. The background vorticity
profile supports two Rossby waves, propagating in the negative (positive)
$x$-direction on the positive (negative) vorticity gradient in $y>0$ ($y<0$).
Viewed individually, the Rossby waves are neutral and propagate against the mean
flow. If, however, they become phase locked, they can interfere constructively,
leading to mutual amplification and hence instability. This is shown
schematically in figure~\ref{fig:CRWmod}, where the two Rossby waves are
represented as perturbed vorticity contours (or equivalently perturbed material
contours) for $y>0$ and $y<0$. The resulting positive and negative vorticity
anomalies are also shown. In this configuration, the transverse flow induced by
each wave acts to amplify the existing transverse material displacement of the
other. There is thus a mutual amplification and instability. This is also
consistent with numerical solutions for instabilities of the flow (\ref{e:tanh})
when $M=F=0$; figure~\ref{fig:tanh_most_unstable_eigenfunction}$a$ shows the
vorticity perturbation of the most unstable mode (with $\alpha=0.44$), in
agreement with figure~\ref{fig:CRWmod}.

We now quantify how free-surface and magnetic effects modify this mechanism for
SWMHD. We use the SWMHD vorticity equation, which is given by
\begin{equation}\label{ch5:CRW1}
	\frac{\mathrm{D} \omega}{\mathrm{D} t} \equiv	
	\frac{\partial \omega}{\partial {}t} + \ub\cdot\grad\omega 
	= -(\grad\cdot\ub)\omega +  M^2 \Bb \cdot \grad{} j + M^2 (\grad\cdot\Bb) j,
\end{equation}
where $\omega$ and $j$ are the $z$-components of the vorticity and electric
current. Using equations~\eqref{ch3:basic-equation3} and
\eqref{ch3:basic-equation4}, \eqref{ch5:CRW1} can be written as
\begin{equation}\label{ch5:CRW2}
	\frac{\mathrm{D} \omega}{\mathrm{D} t} = 
	F^2 \omega \frac{\mathrm{D} h}{\mathrm{D} t} 
	+ F^2 h \omega \grad \cdot \ub 
	+ M^2 (1-F^2 j) \Bb \cdot \grad{} j - M^2 h j \grad \cdot{} \Bb.
\end{equation}
On linearising about the basic state $\Ub_{0} = U(y)\eb_{x}$, $\Bb_{0}
=\eb_{x}$, taking modal solutions of the form~\eqref{ch3:expform}, and noting
that $v=(\dy/\dy{}t + U \dy/\dy{}x)\eta$, where $\eta$ is the cross-stream
displacement, we obtain the the vorticity budget:
\begin{equation}\label{ch5:CRW4}
	\omega = - \eta \Omega' + F^2 h\Omega 
	+ M^2 \frac{j}{U-c},
\end{equation}
where $\Omega = -U'$ is the basic state vorticity. The three contributions to
$\omega$ arise from the advection of the background vorticity and from
shallow-water and magnetic effects.

\begin{figure}
	\begin{center}
	\includegraphics[width=\textwidth]{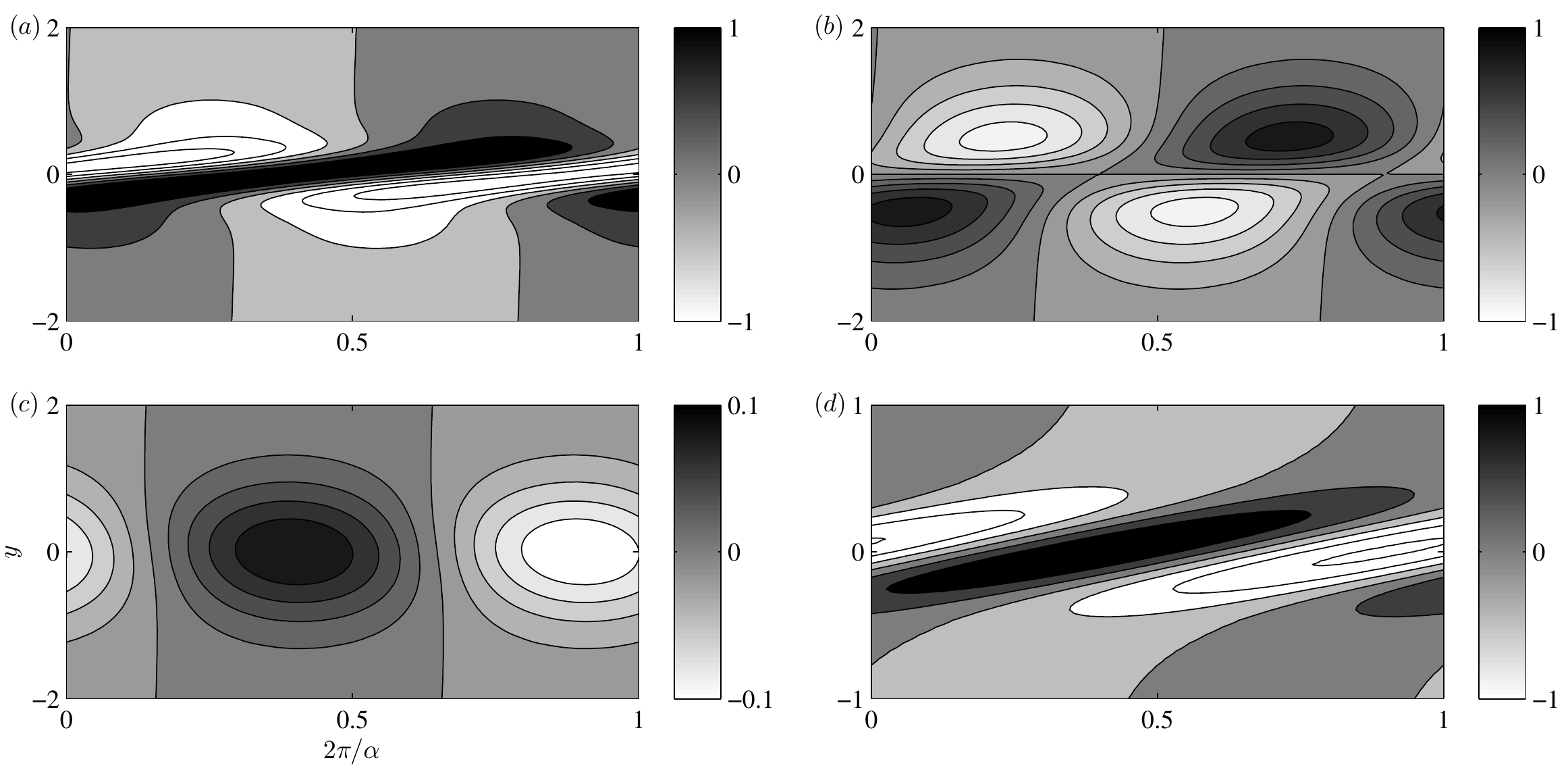}
	\caption{A generic vorticity budget breakdown for the case where neither $F$
	nor $M$ is zero; shown here is $F=0.5$ and $M=0.25$, with ($a$) $\omega$,
	($b$) $-\eta\Omega'$, ($c$) $F^2 h \Omega$, ($d$) $M^2 j/(U-c)$. Notice
	that the vorticity contribution from the magnetic term is substantially
	larger than that from the shallow-water term.}
	\label{fig:omega_contribution_most_unstable_M0.25_F0.5}
	\end{center}
\end{figure}

Inspection of figure~\ref{fig:tanh_contour_FM} shows that the instability is
most vigorous when $M=F=0$; we therefore expect that the vorticity anomalies
from the magnetic and shallow-water effects will be stabilising. The vorticity
$\omega$ and the decomposition~(\ref{ch5:CRW4}) are shown in
figure~\ref{fig:omega_contribution_most_unstable_M0.25_F0.5} for a mode at
$M=0.25$, $F=0.5$. Even though this eigenfunction results from a calculation
with non-zero $F$ and $M$, the $-\eta \Omega'$ contribution has the same
structure as that of figure~\ref{fig:tanh_most_unstable_eigenfunction}$a$. The
extra contributions from non-zero $F$ and $M$ are shown in
figure~\ref{fig:omega_contribution_most_unstable_M0.25_F0.5}$c,d$; both of
these terms are maximised at $y=0$, where they are approximately in phase. Thus,
at the simplest level, they lead to the vorticity anomalies shown by the dashed
circles in figure~\ref{fig:CRWmod}. The transverse flow induced by these
vorticity anomalies counteracts the mutually amplifying transverse flow of the
Rossby waves, and is thus stabilising.

As a further verification of these ideas, we have also adopted a perturbative
approach to the analysis of expression~(\ref{ch5:CRW4}), approximating the
shallow-water and magnetic contributions using the eigenfunction for $M=F=0$. It
can be seen that calculating $F^2 h \Omega$ using $h$ in
figure~\ref{fig:tanh_most_unstable_eigenfunction}$b$ is consistent with the full
linear equations
(figure~\ref{fig:omega_contribution_most_unstable_M0.25_F0.5}$c$). To obtain an
estimate of the magnetic contribution, it is necessary to calculate $j$ using
the governing equations~(\ref{ch3:linear-equ1}) with the velocity obtained when
$M=F=0$. This is slightly more involved than for $h$, but can be shown to
provide a vorticity contribution consistent with
figure~\ref{fig:omega_contribution_most_unstable_M0.25_F0.5}$d$ (see
\citealt{Mak-thesis}).


\subsection{Bickley jet}

In this subsection we consider the basic state velocity defined by
\begin{equation}
	U(y)=\textnormal{sech}^2 y ,
\end{equation}
again with $B(y) \equiv 1$. From inequality~\eqref{thm:hoiland}, the growth rate
$\alpha{}c_i$ is bounded above by $|U'|_{\mathrm{max}}/2=2/(3\sqrt3)$;
furthermore, from stability criterion \eqref{thm:lin-stable2}, this flow is
stable when $M\geq1/2$. When $M=F=0$, even and odd modes are unstable only in
the respective bandwidths $0 < \alpha < 2$ and $0 < \alpha < 1$
\citep[e.g.,][\S31.9]{DrazinReid-Stability}.

\begin{figure}
	\begin{center}
	\includegraphics[width=140mm]{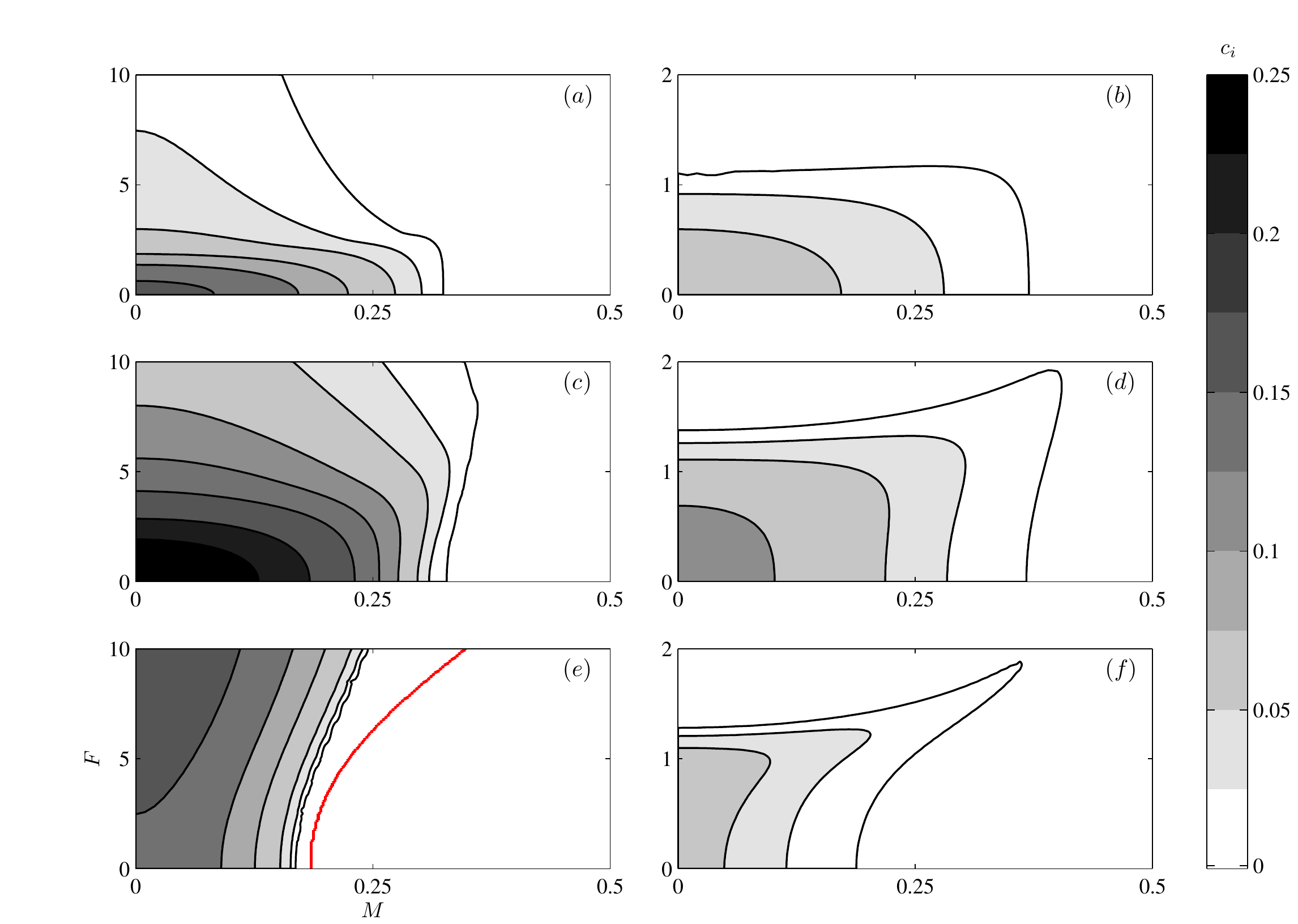}
	\caption{Contours of $c_i$ over $(M,F)$ space at selected $\alpha$ for the
	even (left column) and odd (right column) modes of the Bickley jet, with
	($a$)~$\alpha=0.9$, ($b$)~$\alpha=0.52$, ($c$)~$\alpha=0.3$,
	($d$)~$\alpha=0.23$, ($e,f$) $\alpha=0.05$. The cutoff from the
	asymptotic result \eqref{ch5:bickley_asymp2} is also plotted in panel ($e$).}
	\label{fig:bickley_contour_FM}
	\end{center}
\end{figure}

Figure~\ref{fig:bickley_contour_FM} shows contours of $c_i$ over $(M,F)$ space
for selected values of $\alpha$. As for figure~\ref{fig:tanh_contour_FM}, the
eigenvalues are calculated using a mode-tracking procedure, starting from the
case of $M=F=0$. The values of $\alpha$ correspond to: (i) the most unstable
mode when $M=F=0$ (panels $a$ and $b$); (ii) the mode with highest $c_i$ when
$M=F=0$ (panels $c$ and $d$); (iii) a long-wave disturbance (panels $e$ and
$f$). In all these cases the magnetic field is stabilising. This effect will be
quantified later via a long-wave asymptotic analysis.

\begin{figure}
	\begin{center}
	\includegraphics[width=\textwidth]{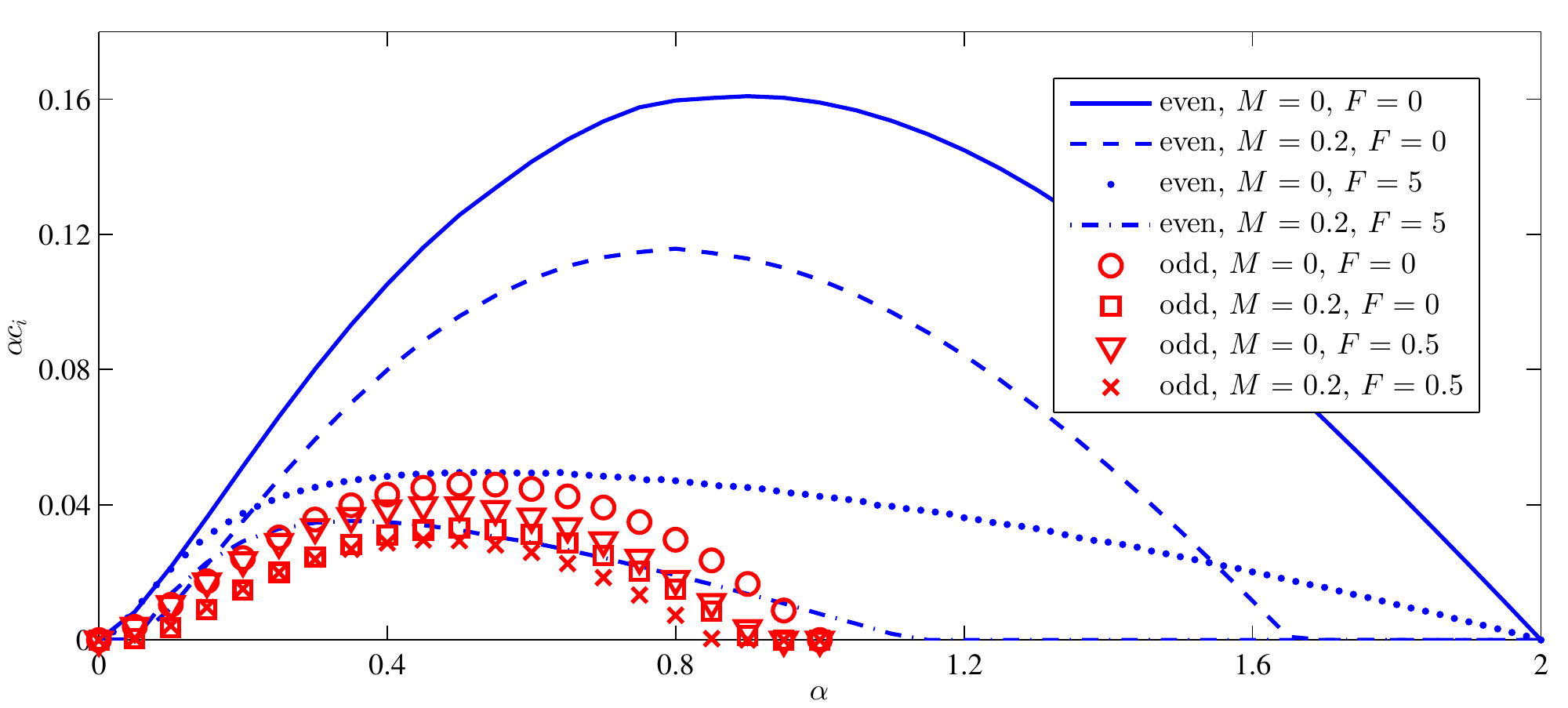}
	\caption{Growth rate of the primary instability of the Bickley jet.}
	\label{fig:bickley_growth_even_odd}
	\end{center}
\end{figure}

Figure~\ref{fig:bickley_growth_even_odd} shows the growth rate at selected
parameter values. In general, the even mode is more unstable than the odd mode.
There are isolated regions where the odd mode is more unstable, although these
do not necessarily correspond to the regions predicted by the stability analysis
for the rectangular jet, described in \S4.

\begin{figure}
	\begin{center}
	\includegraphics[width=\textwidth]{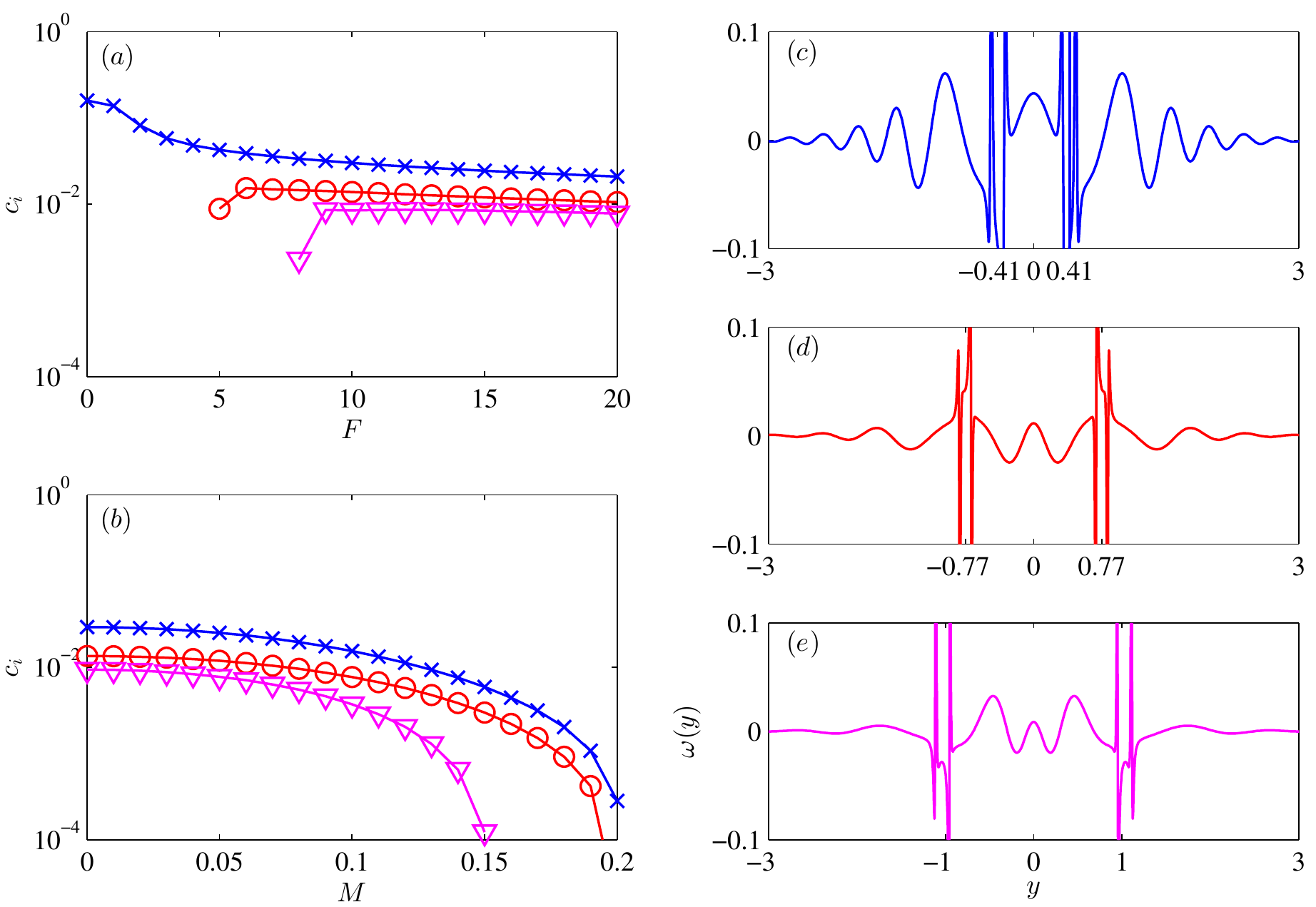}
	\caption{Even mode of the Bickley jet at $\alpha=1$. ($a$) $c_i$ as a function
	of $F$ at $M=0$, $\alpha=1$. ($b$)~ $c_i$ as a function of $M$ at $F=20$.
	Crosses show the primary mode, circles and triangles the first two secondary
	modes. ($c,d,e$)~Cross-sections of the vorticity perturbation for $M=0.05$,
	$F=20$ for the primary and first two secondary modes. The locations where
	$U=c_r$ are labelled on the axes. The spikes reach an amplitude of
	approximately $0.5$; the display range has been reduced in order that the
	oscillations within the core are clearly visible.}
	\label{fig:Bickley_gill_mode_F20_alp1}
	\end{center}
\end{figure}

A natural question to ask, prompted by the findings for the shear layer as well
as the results for the rectangular jet in \S4.2.5, is whether there are
additional modes of instability to those shown in
figure~\ref{fig:bickley_growth_even_odd}. We have performed a scan over $(M, F)$
space at various values of $\alpha$, with randomly generated initial guesses for
$c$ within the smallest rectangle containing the
semi-circle~\eqref{thm:semicircle2}. A substantial number of computations ($20$
different initialisations at over $200$ different parameter values) were carried
out, solving the governing eigenvalue equation with no parity imposed. Other
solution branches were found for large $F$ when $\alpha$ is sufficiently large,
analogous to the secondary modes for the rectangular jet in \S4.2.5. Sample
plots of $c_i$ and cross-sections of the eigenfunctions for the even modes are
shown in figure~\ref{fig:Bickley_gill_mode_F20_alp1}; the results for the odd
mode are similar. As in \S4.2.5, further secondary modes appear as $F$ increases
(figure~\ref{fig:Bickley_gill_mode_F20_alp1}$a$) and the associated growth
rates are small (figures~\ref{fig:Bickley_gill_mode_F20_alp1}$a,b$). However,
in contrast to the results shown in figure~\ref{fig:gill} for the rectangular
jet, we did not find branch crossings in $c_i$ for the Bickley jet; this may
occur at higher values of $F$, but computations are demanding since $c_i$
becomes increasingly small.

There are two distinctive features of the spatial structure of the
eigenfunctions, as shown in
figures~\ref{fig:Bickley_gill_mode_F20_alp1}$c,d,e$. First, there is a core
region consisting of approximately quantised oscillations: the primary mode has
one oscillation, the $n^{\textnormal{th}}$ secondary mode has $n+1$
oscillations. Second, the boundary of the core region is approximately located
where $U(y)=c_r$. Here the eigenfunctions are highly oscillatory and of larger
amplitude, although they are well-resolved in
figures~\ref{fig:Bickley_gill_mode_F20_alp1}$c,d,e$.

\subsection{Long-wave asymptotics for unbounded smooth profiles}\label{sec:LWA}

Further understanding of shear flow instabilities in SWMHD can be obtained by
generalising the long-wave asymptotic procedure of \cite{DrazinHoward62}, who
considered two-dimensional hydrodynamics. The idea is that, for long-wave
disturbances, the leading order behaviour of the instability is determined by
$U(y)$ as $|y| \to \infty$, with higher order corrections determined by the flow
at finite $y$. In this subsection we extend the formalism of
\cite{DrazinHoward62} to SWMHD, but with a uniform magnetic field and no
topography.

We consider the governing equation~\eqref{ch5:numerical_1}, written as
\begin{equation}
\label{ch5:long1}
	Z^2(G''-\alpha^2 K^2 G)+(Z^2)'G'=0,\qquad Z^2 = \frac{S^2}{K^2}.
\end{equation}
We assume that $U_\pm=U(\pm\infty)$ are well defined and that $U'$ (and so
$(Z^2)'$) decays sufficiently rapidly as $|y| \to \infty$. Then, on choosing an
appropriate frame of reference and suitable normalisation for the basic flow,
any velocity profile may be designated as either a \emph{shear layer} if
$U_\pm=\pm1$, or as a \emph{jet} if $U_\pm=0$. 

Adopting the same notation as \cite{DrazinHoward62}, we consider solutions to
\eqref{ch5:long1} of the form
\begin{equation}\label{ch5:long2}
	G(y)=\begin{cases}
	G_{+}(y) = \chi(y)    \exp(-\alpha{K_{+}}y),\ &y>0,\\
	G_{-}(y) =  \theta(y) \exp(+\alpha{K_{-}}y),\ &y<0,\\
	\end{cases}
\end{equation}
where $K^2_{\pm} = 1 -F^2 S^2_{\pm} = 1-F^2 \left( (U_{\pm}-c)^2 - M^2 \right)$.
The perturbations must decay as $|y| \to \infty$; hence $\Real(K_{\pm})>0$. We
consider expansions of the form
\begin{equation}\label{ch5:long3}
	\chi (y) = \sum^{\infty}_{n=0}(+\alpha)^{n}\chi_{n}(y), \qquad 
         \theta (y) =\sum^{\infty}_{n=0}(-\alpha)^{n}\theta_{n} (y),
\end{equation}
with $\chi_0, \theta_0 \to \mathrm{constant} \ (\ne 0)$ and $\chi_n, \theta_n
\to 0$ as $|y| \to \infty$ $(n \ge 1)$. It turns out to be most convenient to
fix $\chi_0(\infty) = \theta_0(-\infty) =1$, and then to accommodate the
necessary degree of freedom in the matching conditions for $G$ at $y=0$, namely
$G_{+}(0)=\Gamma{}G_{-}(0)$ and $G_{+}'(0)=\Gamma{}G_{-}'(0)$ for some constant
$\Gamma$. Consistency thus implies
\begin{equation}\label{eq:matching_condition}
  G_{+}(0) G_{-}'(0) = G_{-}(0) G_{+}'(0).
\end{equation}

Without loss of generality, we shall focus on the equations for $\chi$; those
for $\theta$ follow in a similar fashion. On substituting \eqref{ch5:long2}
(with \eqref{ch5:long3}) into \eqref{ch5:long1}, equating the coefficients at
each order of $\alpha$ gives
\begin{subequations}\label{ch5:long4}
\begin{align}
	0 &= \left( Z^2 \chi_0' \right)', \label{ch5:long4a} \\
	0 &= \left( Z^2 \chi_1' \right)'-K_
	+ \left( 2Z^2 \chi_0'+(Z^2 )'\chi_0 \right), 
	\label{ch5:long4b} \\
	0 &= \left( Z^2 \chi_{n+2}' \right)'-
	K_+ \left( 2Z^2 \chi_{n+1}' + (Z^2)' \chi_{n+1} \right) 
	+ Z^2 (K_+^2 - K^2) \chi_n ,\ \ n\geq0.
	\label{ch5:long4c} 
\end{align}
\end{subequations}
Equation \eqref{ch5:long4a} integrates to $Z^2 \chi_0' = C$, with the conditions
at infinity then giving $C=0$. Thus $\chi_0 = \textrm{constant} = 1$ through our
choice of $\chi_0(\infty)$. Integration of equations (\ref{ch5:long4}$b,c$)
then gives, after some algebra,
\begin{equation}\begin{aligned}\label{ch5:long5}
	&\chi_1 = \int_{\infty}^y \left(1-\frac{Z_+^2}{Z^2}\right)\, \mathrm{d}y_{1},\\
	&\chi_2=\int_{\infty}^{y}\left( \frac{1}{Z^2}
	\int_{\infty}^{y_{1}}(S^2 - S^2_{+})
	\	\mathrm{d}y_2 
	+ K_{+}^2\int_{\infty}^{y_{1}}\left(1-\frac{Z_{+}^2}{Z^2}\right)
	\, \mathrm{d}y_2\right)\, \mathrm{d}y_{1}.
\end{aligned}\end{equation}
The matching condition~\eqref{eq:matching_condition} then leads to the result
\begin{equation}\begin{aligned}\label{ch5:long7}
	0 & = \left( \frac{S_+^2}{K_+}+\frac{S_-^2}{K_-} \right) \\
	  &+ \alpha \left( \int_0^{\infty}(S^2 - S_+^2)\, \mathrm{d}y
	  +\int_{-\infty}^0 (S^2 -S_-^2)\, \mathrm{d}y\right.\\
	&\left.-\frac{S^2_+}{K_+ K_-}\int_{-\infty}^0
	\left(1-\frac{S_-^2}{S^2} \right)\mathrm{d}y
	- \frac{S^2_-}{K_+K_-}\int_0^{\infty}\left(1 - \frac{S_+^2}{S^2} \right)
	\, \mathrm{d}y \right) + O(\alpha^2).
\end{aligned}\end{equation}
On expressing the eigenvalue $c$ as $c= c^{(0)} + \alpha c^{(1)} + \alpha^2
c^{(2)} + \cdots$, equation~\eqref{ch5:long7} then determines the successive
$c^{(j)}$.

Although we have focused on the case of a uniform magnetic field, it is possible
to include a non-uniform field, subject to imposing conditions analogous to
those for $U(y)$. With underlying topography, other assumptions are required
\citep{CollingsGrimshaw80}.


\subsubsection{Shear layers}

For a shear layer, $U_\pm=\pm1$, and the leading order term of
expression~\eqref{ch5:long7} gives
\begin{equation}\label{ch5:long_vortex}
	\frac{(1-c^{(0)})^2-M^2}{\sqrt{1-F^2 \left( (1-c^{(0)})^2-M^2 \right)}}
	+\frac{(1+c^{(0)})^2 - M^2}{\sqrt{1-F^2 \left( (1+c^{(0)})^2-M^2 \right)}}=0.
\end{equation}
This is exactly the eigenvalue equation of the vortex sheet \eqref{ch4:vortex4};
hence, for any shear layer, $c\rightarrow c_{\rm v}$, as defined in
\eqref{ch4:vortex7}, as $\alpha\rightarrow0$. This is not surprising;
sufficiently long waves see the shear layer as a discontinuity in the flow.

When $F$ is sufficiently large, $c_{\rm v}$ is real (see \S4.1). In
this case, following \cite{Blumen-et-al75}, we calculate $c^{(1)}$ to seek a
secondary mode of instability. For the particular case $U(y) = \tanh y$, and
after considerable algebra, we obtain
\begin{equation}\label{ch5:long_c1_outer}\begin{aligned}
	c^{(1)} = 
	&\zi\frac{\sqrt{F^2
	\left((1-c_{\rm v})^2 - M^2 \right)-1}}{4c_{\rm v}^2}
	\frac{(1+c_{\rm v})^2-M^2}{\sqrt{1+4F^2+4F^4M^2}}\times\\
	&\qquad\left(1+\frac{c_{\rm v}^2}{2} 
	\left( \log \left(
	\frac{(1+c_{\rm v})^2-M^2}{(1-c_{\rm v})^2-M^2} \right)
	-\delta_1 - \delta_2 \right) \right.\\
	&\qquad\qquad\left. + \frac{1 -M^2-c_{\rm v}^2}{4M}
	\left[ \log\left(
	\frac{(1+M)^2-c_{\rm v}^2}{(1-M)^2-c_{\rm v}^2}\right)+
	\delta_{1}-\delta_2\right] \right),
\end{aligned}\end{equation}
where
\begin{equation*}
	\delta_{1}=\begin{cases}
	  \pi\zi, &|c_{\rm v}-M|\leq1,\\
	  0, &\textnormal{otherwise},
	  \end{cases}
	\qquad
	\delta_2=\begin{cases}
	  \pi\zi, &c_{\rm v}+M\leq1,\\
	  0, &\textnormal{otherwise}.
	  \end{cases}
\end{equation*}
It may be shown that for $M\rightarrow0$, the expression inside the square
brackets is equal to $4M(1-c_{\rm v})^{-2} + O(M^2)$, and that
equation~\eqref{ch5:long_c1_outer} is equivalent to equation~(21) in
\cite{Blumen-et-al75}. Equation~\eqref{ch5:long_c1_outer} does indeed describe
instability beyond the vortex sheet cutoff. By expanding $c_{\rm v}$ up to
powers of $F^{-4}$, it may be shown from \eqref{ch5:long_c1_outer} that
$\Imag(c^{(1)}) \downarrow 0$ as $F \rightarrow \infty$, so there is no cutoff
at finite $F$.

The analysis leading to expression~\eqref{ch5:long_c1_outer} is valid only when
$c_{\rm v}$ is not small, i.e.\ for $F^2$ not close to $2(1-M^2)^{-1}$ or $M$
not close to $1$. When $2-F^2(1-M^2)^{-1}$ is small, $O(\alpha^{2/3})$ to be
precise, we have $c_{\rm v} \sim \alpha{}c^{(1)} \sim \alpha^{1/3}$. Rescaling
and choosing the appropriate branch so that $\Real(\sqrt{\cdots})>0$, gives
$c^{(1)}$ as the solution of the cubic equation
\begin{equation}\label{ch5:long_c1_inner}
	\frac{6 + 2M^2}{(1-M^2)^3}\left( c^{(1)}\right)^3 
	+ \left(\frac{2}{1-M^2}-F^2\right) c^{(1)} 
	- \zi \alpha \left( \frac{1}{1-M^2} 
	+ \frac{1}{2M}\log\frac{1+M}{1-M} \right) = 0,
\end{equation}
for $\alpha^{2/3}\sim2(1-M^2)^{-1}-F^2\rightarrow0$. When $M=0$,
equation~\eqref{ch5:long_c1_inner} reduces to equation~(23) in
\cite{Blumen-et-al75}. There are two admissible roots with positive imaginary
parts. There is a transition to non-zero real parts when
\begin{equation}\label{ch5:long_c1_cusp}
	F^2_{\textnormal{cusp}}=\frac{2}{1-M^2} 
	- 3\left( \frac{\alpha^2}{4}\frac{6+2M^2}{(1-M^2)^{3}}
	\left( \frac{1}{1-M^2} + \frac{1}{2M}\log\frac{1+M}{1-M} \right)^2
	\right)^{1/3}.
\end{equation}
This expression reduces to $F^2_{\textnormal{cusp}} = 2 - 3(6\alpha^2)^{1/3}$
when $M \to 0$, as given by \cite{Blumen-et-al75}.

\begin{figure}
	\begin{center}
	\includegraphics[width=\textwidth]{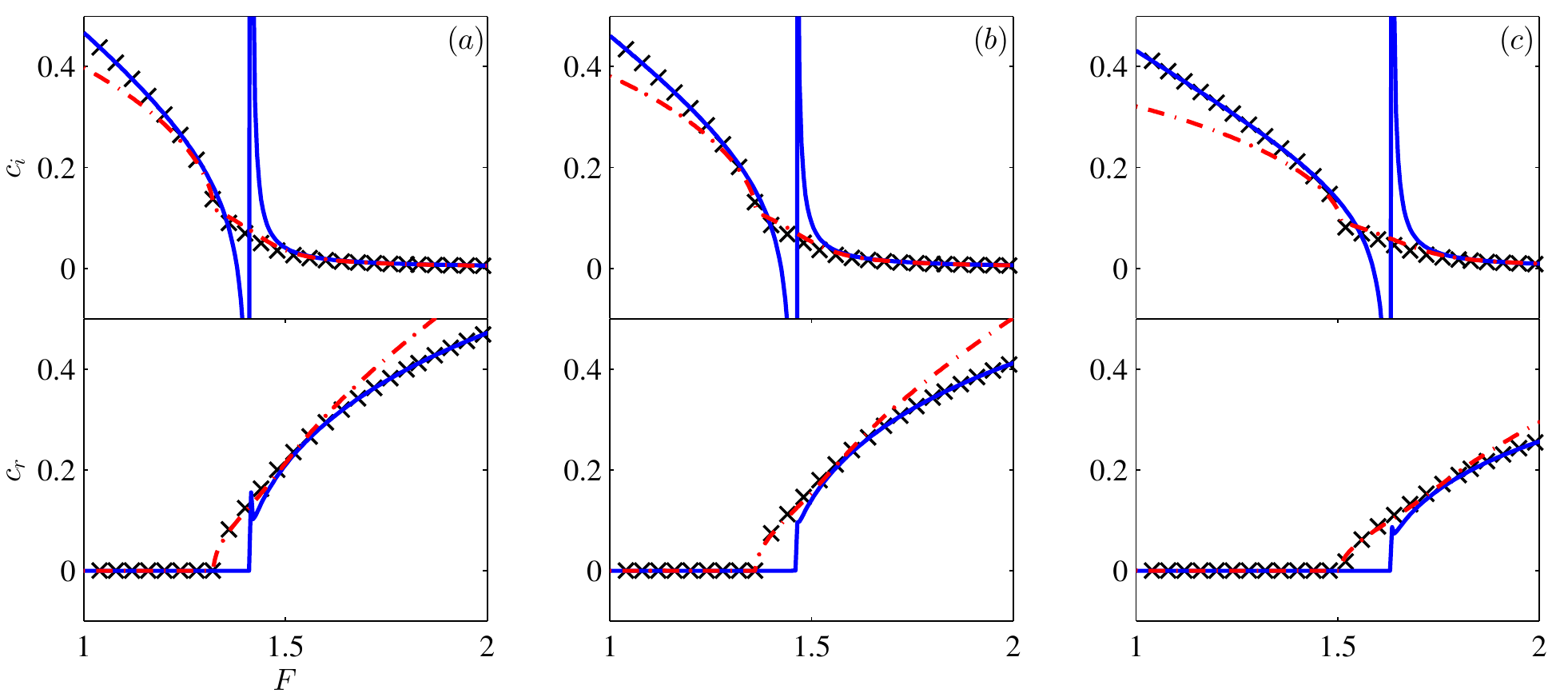}
	\caption{$c_i$ and $c_r$ versus $F$ at $\alpha=0.01$ for ($a$) $M=0$, ($b$)
	$M=0.25$, ($c$) $M=0.5$. The crosses are the computed results for $U(y)=\tanh
	y$, the solid lines are the asymptotic result $c_{\rm v} + \alpha{} c^{(1)}$
	with $c^{(1)}$ given by \eqref{ch5:long_c1_outer}, and the dot-dashed lines
	are the inner expansion given by the relevant solution to the cubic
	equation~\eqref{ch5:long_c1_inner}.}
	\label{fig:second_mode_asymp_alp001}
	\end{center}
\end{figure}

The asymptotic results, together with the numerical computations of \S5.2, are
presented in figure~\ref{fig:second_mode_asymp_alp001}. The agreement between
the two, including the location of the cusp given by
equation~\eqref{ch5:long_c1_cusp}, is excellent.

\subsubsection{Jets}

For a jet, $U_{\pm}=0$ and \eqref{ch5:long7} simplifies to
\begin{equation}\label{ch5:long_jet}
	0=\frac{2S_0^2}{K_0} + \alpha\left( \int_{-\infty}^{\infty}(S^2 - S_0^2)
	\, \mathrm{d}y
	- \frac{S_0^2}{K_0^2} \int_{-\infty}^{\infty}
	\left(1 - \frac{S_0^2}{S^2} \right)\, \mathrm{d}y \right) + O(\alpha^2),
\end{equation}
where $S_0^2 = (0-c)^2 - M^2$. Here, for a fixed value of $F$, we need to
consider different regimes for $M$.

For $F^2=O(1)$, if $M^2 = O(1)$ then $c^{(0)}$ and $c^{(1)}$ are real. To find
an instability we need to consider the regime $M^2 \sim \alpha$, which implies
$c^{(0)} = 0$. At the next order, we choose to balance the first two terms on
the right hand side of \eqref{ch5:long_jet}, assuming that the second integral
is negligible; this is confirmed by the analysis of Appendix~\ref{appendix1}.
Defining $E=\int^{+\infty}_{-\infty}U^2/2\ \textnormal{d}y$, where $U$ is
assumed to decay sufficiently rapidly that $E$ is finite, we obtain
\begin{equation}\label{ch5:long_jet1}
	c\sim{} \zi \sqrt{\alpha{}E - M^2} \qquad \textnormal{as} \qquad
	M^2 \sim \alpha \to 0,\quad F^2 = O(1).
\end{equation}
The corresponding result for compressible hydrodynamics was derived by
\cite{GillDrazin65}, and for incompressible MHD by \cite{Gedzelman73}.

For large $F$, the regime of interest is $M^2 \sim \alpha$, $F^2 \sim
\alpha^{-1}$. Considering the same balance as above gives
\begin{equation}\label{ch5:long_jet2}
	c \sim{} \zi \left( \frac{\alpha^2 F^2 E^2}{2} - M^2 +
	\sqrt{\alpha^2 E^2 + \frac{\alpha^4 F^4 E^4 }{4}} \right)^{1/2} \qquad
	\textnormal{as} \qquad F^{-2} \sim{} M^2 \sim \alpha \to 0.
\end{equation}
This result reduces to \eqref{ch5:long_jet1} in the limit of small $F$.
Equations \eqref{ch5:long_jet1} and \eqref{ch5:long_jet2} are the extensions to
smooth velocity profiles of equations \eqref{ch4:jet_even_asymp2} and
\eqref{ch4:jet_even_asymp3}, which are the dispersion relations for the even
mode of the top-hat jet. It is interesting to note that the odd modes are not
recovered by this analysis.

\begin{figure}
	\begin{center}
	\includegraphics[width=\textwidth]{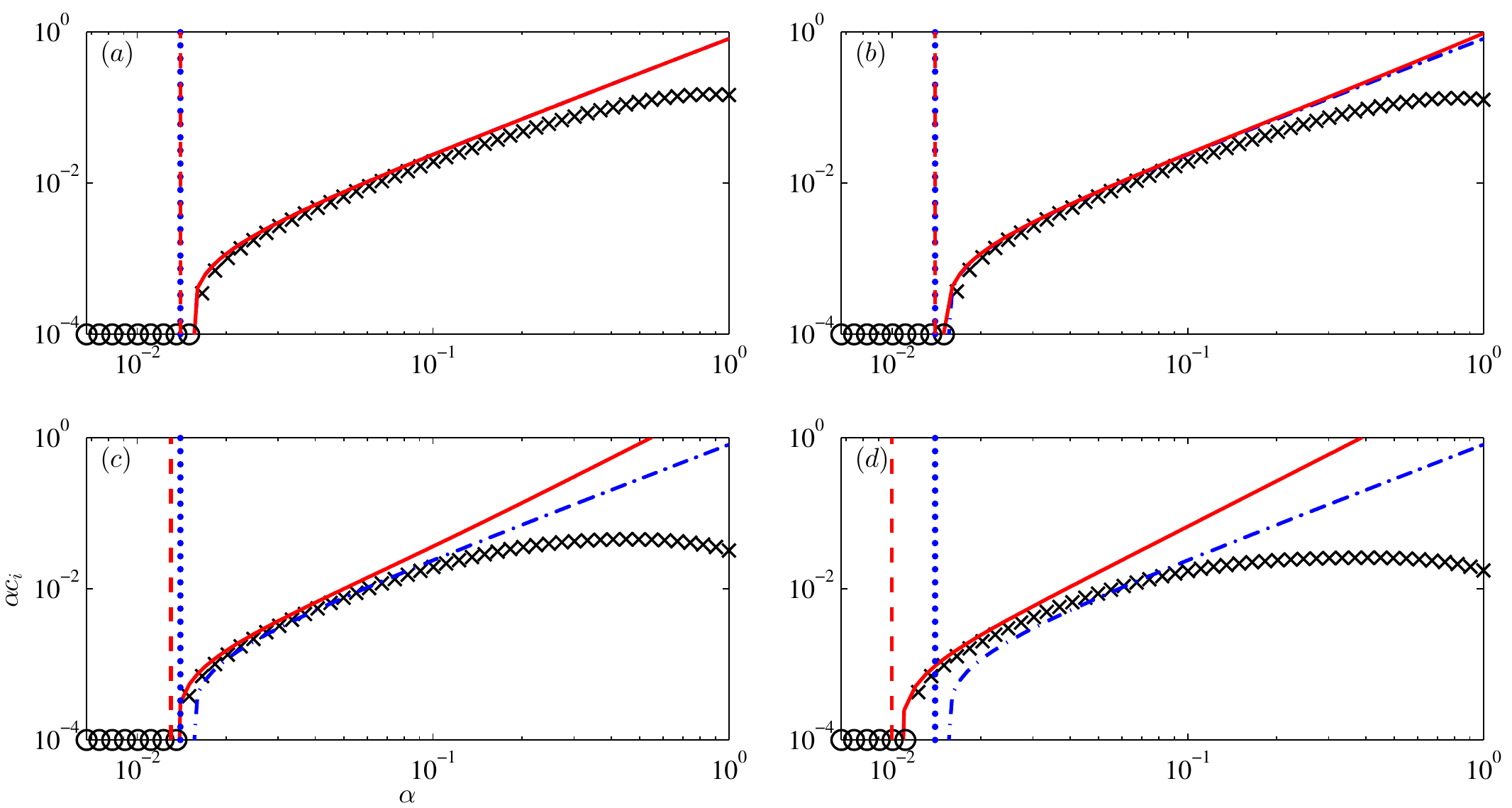}
	\caption{Comparison of the computed growth rates for the even modes (crosses)
	and the predicted growth rates from the asymptotic results for $U(y) =
	\textnormal{sech}^2 y$, for $M=0.1$ and ($a$) $F=0$, ($b$) $F=1$, ($c$) $F=5$,
	($d$) $F=10$. $\alpha \times \textrm{Im}$\eqref{ch5:bickley_asymp1} is given
	by the dot-dashed line (cutoff plotted as vertical dotted line) and $\alpha
	\times \textrm{Im}$\eqref{ch5:bickley_asymp2} is given by the solid line
	(cutoff plotted as vertical dashed line).}
	\label{fig:bickley_small_alp_asymp_B01}
	\end{center}
\end{figure}

For the Bickley jet, $2E = \int^{+\infty}_{-\infty}(\textnormal{sech}^2y)^2\
\textnormal{d}y=4/3$ and \eqref{ch5:long_jet1} and \eqref{ch5:long_jet2} become
\begin{equation}
\label{ch5:bickley_asymp1}
	c \sim \zi \sqrt{\frac{2}{3} \alpha - M^2} \qquad \textnormal{as} \qquad
	M^2 \sim \alpha \to 0, \quad F^2 = O(1) ,
\end{equation}
and
\begin{equation}
\label{ch5:bickley_asymp2}
	c \sim \zi \left( \frac{2}{9} \alpha^2 F^2 - M^2 
	+ \frac{2}{3} \sqrt{\alpha^2 + \frac{\alpha^4 F^4}{9}} \right)^{1/2} 
	\qquad \textnormal{as} \qquad 
	F^{-2} \sim{} M^2 \sim \alpha \to 0.
\end{equation}
In figure~\ref{fig:bickley_small_alp_asymp_B01} the growth rates given by
\eqref{ch5:bickley_asymp1} and \eqref{ch5:bickley_asymp2} are plotted against
those determined numerically from equation~\eqref{ch5:numerical_1}; there is
good agreement at small $\alpha$. The cutoff implied by
\eqref{ch5:bickley_asymp2} is also shown in
figure~\ref{fig:bickley_contour_FM}$e$.

The above asymptotic procedure does not yield the odd modes for smooth velocity
profiles; for these modes, $c_r$ remains $O(1)$ as $\alpha \to 0$, as can be
verified numerically. The difficulty can be traced back to
equation~\eqref{ch5:long1} since, when $M^2 \sim \alpha$, $Z^2$ becomes small
when $U \approx c_r$. The standard asymptotic procedure leading to
equations~\eqref{ch5:long4} breaks down. Instead it is found that $c_i \sim
\alpha^{2/3}$, a result that has been derived for the hydrodynamic case by
\cite{DrazinHoward62}.



\section{Conclusions and discussions}

The SWMHD equations, introduced by \cite{Gilman00}, are a useful model for
studying MHD in thin stratified fluid layers. We have investigated the linear
instability of parallel shear flows with an aligned magnetic field in planar
geometry with no background rotation. The instability of hydrodynamic shear
flows is a classical problem, with well known results such as Rayleigh's
inflection point criterion, Howard's semi-circle theorem and H{\o}iland's growth
rate bound, supplemented by extensive asymptotic and numerical results for
various idealised flows. Motivated by geophysical and astrophysical
considerations, previous authors have extended separately the analysis to
include the influence of shallow-water dynamics and a magnetic field. Here, for
the first time, we have applied this classical approach to the SWMHD system,
complementing previous work on the instability of specific flow configurations
in spherical geometry \citep{GilmanDikpati02, Dikpati-et-al03}.

We first considered the stability of arbitrary flow and field profiles, leading
to the growth rate bound~\eqref{thm:hoiland} and the semi-circle
theorems~\eqref{thm:semicircle1} and \eqref{thm:semicircle2}, which, in turn,
imply the stability criteria~\eqref{thm:lin-stable1} -- \eqref{thm:lin-stable3}.
More detailed investigations focused on shear layers and jets, which are known
from hydrodynamical problems to be canonical flows \citep{DrazinHoward62}. For
each of these we considered both discontinuous profiles, for which dispersion
relations can be obtained analytically, and analogous smooth profiles, for which
numerical solutions were obtained. We also took the background magnetic field to
be uniform; this simple choice already leads to interesting stability
characteristics. The imposed field strength is measured by a parameter $M$,
whilst the stratification is measured by the Froude number $F$.

For the shear layer we considered the vortex sheet and the hyperbolic tangent
velocity profile. For both profiles, a key finding is that, although increasing
$M$ or $F$ in the absence of the other is seen to be stabilising, their combined
effects can offset each other, resulting in a tongue of instability for
arbitrarily large $F$ and $M$ approaching unity. However, the strongest
instabilities are found at smaller $F$ and $M$. For the smooth shear layer,
these were interpreted through the counter-propagating Rossby waves (CRWs)
mechanism, which has been used to understand hydrodynamic shear instabilities
\citep[e.g.,][]{Carpenter-et-al13}. Here we show that the vorticity anomalies
associated with shallow-water effects and magnetic tension oppose those for the
underlying CRW mechanism. As far as we can tell, this is the first time such a
dynamic argument has been invoked to explain the stabilisation of shear flow
instabilities by a magnetic field. For the smooth shear layer alone, there is a
weak secondary instability when $F>1$, analogous to those found in other
hydrodynamic systems \citep{Blumen-et-al75, Balmforth99}. Further insight was
obtained by performing a long wave asymptotic analysis for an arbitrary shear
profile, extending the hydrodynamic treatment of \cite{DrazinHoward62} to SWMHD.
There is a primary mode of instability, yielding a dispersion relation identical
to that of the vortex sheet, thus showing the widespread applicability of the
vortex sheet model. For sufficiently large $F$, the long wave analysis recovers
the weaker secondary mode.

For jets, we considered the top-hat velocity profile and, as its smooth
counterpart, the Bickley jet. For both profiles, the modes of instability are
either even or odd about the jet axis. At large $\alpha$, dynamics on the flanks
of the top-hat become uncoupled, and hence the instabilities become those of the
vortex sheets on the flanks. However, this behaviour does not carry over to the
Bickley jet, since the even and odd modes are stabilised when $\alpha>2$ and
$\alpha>1$, respectively. For small $\alpha$, the even mode is only unstable
when $M<M_c=O(\alpha^{1/2})$. This result applies for both the top-hat profile
and arbitrary smooth jets, showing the widespread applicability of the top-hat
model for the even mode at small $\alpha$. For small $\alpha$, the odd mode is
also only unstable for sufficiently small $M$. The growth rate $c_i =
O(\alpha^{1/2})$ for the top-hat profile, but $c_i = O(\alpha^{2/3})$ for the
Bickley jet, so the analogy between the top-hat and smooth velocity profiles is
lost. Although we can be precise about the behaviour at large and small
$\alpha$, in general either the even or odd mode may
be more unstable. Finally, in addition to these two primary modes of
instability, there exist multiple secondary modes of instability at large $F$,
with relatively small growth rates.

Given the motivation for this study, it is important to consider the
implications of our results for the solar tachocline. Adopting the tachocline
radius ($\approx 0.7R_\odot$) as a characteristic length scale $L_0$ and taking
$U_{0}=500 \,{\rm m \, s^{-1}}$ as a characteristic velocity, as discussed in
\S2, the $e$-folding time for an instability is given by $\hat{t}=(\alpha
c_i)^{-1}(L_{0}/U_{0}) \approx 10^6 (\alpha c_i)^{-1} s$. With $F \approx 0.1$,
$\alpha c_i \approx 0.1$ over the range of $M$ anticipated for the tachocline,
for both shear layers and jets (see figures~\ref{fig:tanh_growth} and
\ref{fig:bickley_growth_even_odd}), leading to large-scale instabilities with
$\hat{t} = O(10^7){\rm s}$, a relevant timescale for solar dynamics.

It is hard to draw direct comparisons with previous results on SWMHD
instabilities \citep{GilmanDikpati02, Dikpati-et-al03, GilmanCally07}. These
studies, which were focused on the tachocline, were in spherical geometry with
background rotation. Furthermore, the main emphasis of the work was on the
destabilisation of hydrodynamically stable velocity profiles by non-uniform
toroidal fields.

Having investigated the quite complicated behaviour that results from our
idealised model, there are, within the present geometry, several natural
extensions of our study. The inclusion of rotation could be considered; even in
the hydrodynamic case this modifies the nature of the instabilities. Non-uniform
magnetic fields open up the possibility of destabilising hydrodynamically stable
velocity profiles, as in \cite{GilmanDikpati02} and \cite{Dikpati-et-al03}. The
nonlinear evolution of the instability and the associated changes in the mean
flow are clearly of interest; some preliminary results are given in
\cite{Mak-thesis}.

This work was supported by the STFC doctoral training grant ST/F006934/1. JM
thanks Eyal Heifetz for helpful discussions.

\appendix

\section{Consistency checks for the long-wavelength jet analysis}
\label{appendix1}

The aim of this appendix is to show that 
\begin{equation}
\label{ch5:consistency}
	\frac{c^2-M^2}{1-F^2(c^2-M^2)}\int^{+\infty}_{-\infty}
	\left( 1-\frac{c^2-M^2}{(U-c)^2-M^2} \right) \ \textnormal{d}y
\end{equation}
is $O\left( \alpha \log (\alpha) \right)$, and hence that the asymptotic
analysis of \S\,5.5.2 is consistent.

Following \cite{DrazinHoward62}, we assume that $|U| \leq{} A \ex^{-a|y|}$,
which is satisfied for the Bickley jet. They make the additional assumption that
$c$ is `almost pure imaginary'; here we adopt the modified assumption that
\begin{equation}\label{ch5:consistency2}
	\frac{|c|^2 + M^2}{c_i^2} \leq{} N = O(1),
\end{equation}
which is supported by both our numerical and asymptotic results.

Consider first the case of $F=0$ for which \eqref{ch5:consistency} is given by
\begin{equation}\label{ch5:consistency3}
	I = (c^2 - M^2) \int^{\infty}_{-\infty} 
	\left( 1 - \frac{c^2 - M^2}{(U-c)^2 - M^2} \right) \ \textnormal{d}y .
\end{equation}
For $y>0$, we split the range of integration into $(0, \lambda)$ and $(\lambda,
\infty)$, where
\begin{equation}\label{ch5:split}
\lambda = \log\left(\frac{A}{c}\right)^{1/a} = O \left( \log(\alpha) \right),
\end{equation}
since $c^2 = O(\alpha)$. Then
\begin{multline}
   \label{ch5:consistency4}
   	(c^2 - M^2) \int^\lambda_0 (\cdots)\ \textnormal{d}y
	\leq|c^2 - M^2| \int^\lambda_0 
	\left( 1+\frac{|c|^2 + M^2}{c_i^2} \right) \ \textnormal{d}y \\
	\leq|c^2 - M^2| \lambda \left( 1+N \right) 
	=O(\alpha\log\alpha),
\end{multline}
where we have used the usual integral inequalities, the inequality
$|(U-c)^2-M^2|\geq{}c_{i}^2$, the assumption that $M^2=O(\alpha)$, and the
derived result~\eqref{ch5:long_jet1} that $c=O(\alpha^{1/2})$.

Similarly,
\begin{multline*}
	(c^2 - M^2) \int^{\infty}_\lambda(\cdots)\ \textnormal{d}y
    = (c^2 - M^2) \int^{\infty}_\lambda 
	\left( \frac{U^2 - 2Uc}{(U-c)^2 - M^2}\right) {\rm d} y \\
      \leq \frac{|c^2 - M^2|}{c_i^2} 
	\int^{\infty}_\lambda \left( U^2 + 2|U||c| \right) \textnormal{d}y
      \leq{} N \left( \dfrac{A^2 \ex^{-2a\lambda}}{2a}
	+\dfrac{2A |c| \ex^{-a \lambda }}{a}\right)
	=O(\alpha).
\end{multline*}
The dominant contribution is from \eqref{ch5:consistency4}, and hence
$I=O(\alpha\log\alpha)$.

When $F$ is non-zero, the only difference is in the pre-factor to the
integral~\eqref{ch5:consistency}. However, with $c$ from either
\eqref{ch5:bickley_asymp1} or \eqref{ch5:bickley_asymp2} the pre-factor is
$O(1)$ and we may conclude that $I=O(\alpha\log\alpha)$.

\newpage
\bibliographystyle{jfm}

\label{lastpage}

\end{document}